\documentclass[pra, aps, twocolumn, superscriptaddress, nofootinbib, tightenlines, nobibnotes]{revtex4-1}

\usepackage{amssymb}
\usepackage{graphicx}
\usepackage{dcolumn}
\usepackage{bm}
\usepackage{hyperref}
\usepackage{xcolor}
\usepackage[caption=false]{subfig}
  \usepackage[utf8]{inputenc}
 \usepackage[english]{babel}
\usepackage{amsmath}
 \usepackage{hyperref}
 \usepackage{amsmath}
 
\definecolor{lapislazuli}{rgb}{0, 0, 1}
\definecolor{YKblue}{rgb}{0.0, 0.18, 0.65}
\definecolor{carmine}{rgb}{0.81, 0.09, 0.03}
\definecolor{lavender}{rgb}{0.84, 0.49, 0.87}

\hypersetup{
colorlinks=true,
linkcolor=lapislazuli,
linktoc=page,
citecolor=lapislazuli,
urlcolor=lapislazuli
}

\usepackage{amssymb}
\usepackage{mathtools}

\everymath{\displaystyle} 
\newcommand{\pr}[1]{\ensuremath{\left[#1\right]}} 
\newcommand{\pc}[1]{\ensuremath{\left(#1\right)}} 
\newcommand{\px}[1]{\ensuremath{\left\lbrace#1\right\rbrace}} 
\newcommand{\av}[1]{\ensuremath{\left\langle#1\right\rangle}} 

\newcommand{\vect}[1]{\boldsymbol{#1}}

\begin{document}

\title{ Quantum dynamics of Bose-polaron in a $d$-dimensional Bose Einstein condensate}

\author{M. Miskeen Khan}
\affiliation{Instituto Superior T\'ecnico, Universidade de Lisboa, Portugal}
\affiliation{ICFO -- Institut de Ciències Fotòniques, The Barcelona Institute
of Science and Technology, Av. Carl Friedrich Gauss 3, 08860 Castelldefels (Barcelona), Spain}
\affiliation{Instituto de Plasmas e Fus\~ao Nuclear, Instituto Superior T\'ecnico, Universidade de Lisboa, Portugal}

\author{H. Ter\c{c}as}
\affiliation{Instituto Superior T\'ecnico, Universidade de Lisboa, Portugal}
\affiliation{Instituto de Plasmas e Fus\~ao Nuclear, Instituto Superior T\'ecnico, Universidade de Lisboa, Portugal}

\author{J. T. Mendon\c{c}a}
\affiliation{Instituto Superior T\'ecnico, Universidade de Lisboa, Portugal}
\affiliation{Instituto de Plasmas e Fus\~ao Nuclear, Instituto Superior T\'ecnico, Universidade de Lisboa, Portugal}

\author{J. Wehr}
\affiliation{Department of Mathematics and Program in Applied Mathematics
University of Arizona
Tucson, AZ 85721-0089
USA}

\author{C. Charalambous}
\affiliation{Instituto de Física Interdisciplinar y Sistemas Complejos IFISC (CSIC-UIB), 07122 Palma de Mallorca, Spain}
\affiliation{ICFO -- Institut de Ciències Fotòniques, The Barcelona Institute
of Science and Technology, Av. Carl Friedrich Gauss 3, 08860 Castelldefels (Barcelona), Spain}

\author{M. Lewenstein}
\affiliation{ICFO -- Institut de Ciències Fotòniques, The Barcelona Institute
of Science and Technology, Av. Carl Friedrich Gauss 3, 08860 Castelldefels (Barcelona), Spain}
\affiliation{ICREA, Pg. Llu\'is Companys 23, 08010 Barcelona, Spain}

\author{M. A. Garcia-March}
\affiliation{ICFO -- Institut de Ciències Fotòniques, The Barcelona Institute
of Science and Technology, Av. Carl Friedrich Gauss 3, 08860 Castelldefels (Barcelona), Spain}
\affiliation{Instituto Universitario de Matem\'atica Pura y Aplicada, Universitat Polit\`ecnica de Val\`encia, E-46022 Val\`encia, Spain}

\begin{abstract}
We study the quantum motion of an impurity atom immersed in a Bose Einstein condensate in arbitrary dimension. The Bogoliubov excitations of  the Bose Einstein condensate act as a bosonic bath for the  impurity. We  present a detailed derivation of the $d$-dimensional Langevin equations that describe the quantum dynamics of the system, and of the associated generalized  tensor that describes the spectral density in the full generality. When the impurity is not trapped, we calculate the mean square displacement, showing that the motion is super diffusive. We obtain also explicit expressions for the super diffusive coefficient in the small and large temperature limits. We find that, in the latter case, the maximal value of this coefficient is the same in all dimensions. We study also the behaviour of the average energy and compare the results for various dimensions. In the trapped case, we study squeezing and  find that the stronger position squeezing can be obtained in lower dimensions. We quantify the non-Markovianity of the particle's motion, and find that it increases with dimensionality. 

\end{abstract}
\maketitle
\section{Introduction $-$}

The concept of a quasiparticle plays a fundamental role in physics, allowing to greatly simplify the description of numerous complex phenomena. A paradigmatic classical problem, in which quasiparticles appear, is the study of an electron interacting with a surrounding dielectric crystal. Its dynamics can be approximated by a much simpler dynamics of an electron with a different mass, called polaron, traveling through free space. This classical theory (see historical note in  \cite{2015Dykman}) keeps inspiring new developments in physics.  In particular, it plays an important role in the recent studies of the  Bose polaron -- the quasiparticle associated with an impurity immersed in a Bose-Einstein condensate (BEC). 

Bose polarons were investigated in diverse experiments on impurities immersed in  bosonic gases.  To begin with, the quantum dynamics of impurities in Bose gases were examined in \cite{Catani2012,Spethmann2012},  while technical  aspects of experiments with Cs impurities  were studied in \cite{2015Hohmann}. The  phononic Lamb shift in the context of ultracold bosons was observed in \cite{Rentrop2016}. In addition, in these first experiments, charged, ionic or fixed impurities and their dynamics  were studied:  a quantum spin of a localized neutral impurity \cite{2018Schmidt}, fermions in a Bose gas  \cite{2013Scelle,2013Balewski}, ions embedded in a BEC \cite{2010Zipkes, 2010Schmid}.  Quantum dynamics of spin impurities and fermions immersed in a Bose gas in an optical lattice were studied in Refs. \cite{2013Fukuhara,2006Ospelkaus}. More recent experiments define the state-of-the-art of the field: in \cite{Jorgensen2016}  existence of a well-defined quasiparticle state of an impurity interacting with a BEC was demonstrated, while in \cite{Hu2016} the strong interacting regime, which is natural for polaron problems, was investigated. In \cite{Yan2019}, a Bose polaron was studied near criticality, which provided important insights into the physics of quasiparticles in the vicinity of quantum critical points, that are otherwise much more difficult to study in other physical systems.   

The polaron theory was first developed in the strong coupling limit, and later extended to the intermediate and weakly interacting regimes. In the context of the Bose polaron problem, a large part of the  theoretical effort deals with  the weak  regime,  described by the so called Fr\"ohlich Hamiltonian. This theoretical approach  studies  effective mass, quantum dynamics,  \cite{Tempere2009,  Shashi2014, Grusdt2014b, Volosniev2015,Grusdt2016, Shchadilova2016b,2020Mysliwy}, collision dynamics \cite{2018Lingua},  the behaviour in a $d$-dimensional BEC near the critical temperature \cite{2018Pastukhov}, and related aspects of the system.   Some studies in the weak regime considered the impurity as a quantum Brownian particle in a BEC or in a so called  Luttinger liquid \cite{2017Lampo, 2018Lampo,2019LampoBook, Cugliandolo2012, Bonart2013}.  Importantly, Monte Carlo studies, in some instances beyond the regime of validity of the Fr\"ochlich Hamiltonian, allow to benchmark the aforementioned theoretical results  \cite{Ardila2015, Ardila2016, Grusdt2017, Ardila2018}.  Other works focused on the intermediate and strong coupling regimes \cite{Cucchietti2006, Rath2013, Benjamin2014, Brunn2015, Grusdt2016Feb, Shchadilova2016, 2018Grusdt, 2019Seiringer,Drescher2020}, and on the non-zero temperature  systems \cite{ Levinsen2017,  Guenther2018, 2019Liu, Dzsotjan2019}. 

Yet other works studied the quenched dynamics and orthogonality catastrophe using the  multi-configuration time-dependent Hartree method, both in weak and strong interaction regimes  \cite{ 2019Mistakidis, 2019Mistakidisc, 2019Mistakidisb}. Also, several papers investigated  ejection and injection spectroscopy related to Bose polarons and the orthogonality catastrophe \cite{Liu2020, Guenther2020}, as well as bound states   \cite{Levinsen2015,2017Sun,Yoshida2017, 2018Shi}. A number of notable works study two polarons immersed in a BEC (Bose bi-polaron)  \cite{2018Camacho-Guardian, 2019Charalambous} and the problem of an impurity in a two-component BEC \cite{2018Ashida,2020Charalambous}. There has recently been a renewed interest in the polaron problem in mathematical physics literature.  See in particular: \cite{DeRoeck2011, mukherjee2018identification, ROBERT2019strong}.
Finally a series of papers deal with applications of Bose polarons in quantum thermometry \cite{2016Hohmann, Correa2017,  Mehboudi2019, Bouton2020} and thermodynamics \cite{2018Miller, 2019Niedenzu, 2019CharalambousNJP}. 
 
In the present paper we approach the Bose polaron problem from the quantum  open systems perspective. This framework has been already used to understand several questions related to Bose polarons \cite{2017Lampo, 2018Lampo,2019LampoBook,Cugliandolo2012, Bonart2013, 2019Charalambous, 2020Charalambous, Mehboudi2019, 2019CharalambousNJP}.   The study of the Bose polaron from this perspective leads to quantum stochastic equations with  inhomogeneous damping and multiplicative noise.  This is not always possible in the case of non-Ohmic spectral densities \cite{2005Barik, 2018Lim, 2020Lim}, presenting a challenge for mathematical physics.  Particular care is required in higher dimensions, where linearization of the inhomogeneous damping and diffusion is more difficult to control.  In this paper, our first goal is to offer a detailed derivation of the quantum Langevin equations and the associated generalized $d$-dimensional spectral density, which describe from an open quantum system perspective the dynamics of an impurity immersed in a BEC.  In one dimension, this was done in detail in \cite{2017Lampo}, and some hints were given for  two and three dimensions. Performing the calculation in two and three dimensions offers information, valuable for many experiments. When the impurity is not trapped, we calculate (in all dimensions)  the mean square displacement and the average energy, together with analytic expressions valid  in different limits, and discuss the effects of dimensionality on the dynamics. For the trapped impurity, we determine the stationary state, and analyze the position squeezing effect by employing the covariance matrix.  We finally study the non-Markovian character of this dynamics in all dimensions.  We perform  our calculations with experimentally feasible parameters, making sure that the parameters used lie within the regimes of validity of this model. An important  contribution of the paper is that it offers a comprehensive understanding of  the quantum dynamics and non-Markovianity in different dimensions and experimentally realistic situations, not discussed in previous investigations. 

The paper is organized as follows: in section \ref{sec:Ham} we present the assumptions and derivations that permit to obtain the (linearized) quantum Brownian motion Hamiltonian (cf. Eq. \eqref{eq:HamiltonianFrohlichIntFinal}) from the initial second quantized one.  In section \ref{sec:SD} we obtain the generalized $d$-dimensional spectral density. We study, in all dimensions, the quantum dynamics in the non-trapped and trapped case in section \ref{sec:dynamics}, and non-Markovianity in section \ref{sec:memory}. We finally conclude in section \ref{sec:conclusions}. In appendices  \ref{App:AA} and \ref{App:BB} we include the detailed derivation of the generalized $d$-dimensional spectral density and   vectorial  quantum Langevin equations, as they are {\it per se} important results of this paper.  In  appendix \ref{App:CC}  we present the expressions for the position and momentum variances of the generalized Langevin equations. Finally,  in  appendix \ref{App:DD} we discuss the the validity of our approximations. 

    \section{Hamiltonian and Bogoliubov Modes$-$}\label{sec:Ham}
We start by considering an impurity atom with mass $m_{\rm I}$  immersed in a $d$-dimensional ultracold gas of $N$ bosons. The interaction between the bosons occurs through the scattering potential $V_{\rm B}(\bf{r})$. We denote by  $\Psi (\bf{r})$ ($\Psi ^{\dagger}(\mathbf{r})$)  the annihilation (creation) field operator of the atoms at the position $\bf{r}$, which fulfills canonical bosonic commutation relations $[\Psi (\mathbf{r}),\Psi ^{\dagger}(\mathbf{r})]=\delta(\bf{r}-\bf{r}{'})$. The bosonic density therefore takes the form $n_{\rm B}=\Psi^{\dagger} (\mathbf{r}) \Psi (\mathbf{r})$. The total Hamiltonian is given by
\begin{align}
H= H_{\rm I}+H_{\rm B}+H_{\rm BB}+H_{\rm IB}.
 \label{eq:Hamiltonian}
 \end{align}
Here, the four terms represent the Hamiltonians of the impurity being kept in an external potential $U_{\rm ext}(\mathbf{r})$,   bosons in a potential $V_{\rm ext}(\mathbf{r})$, the boson-boson atomic interaction and the impurity-boson atomic interaction, respectively. Within the second quantization formalism,  their explicit forms are \cite{2017Lampo}
\begin{align}
&H_{\rm I}=\frac{\mathbf{P}^2}{2m_{\rm I}}+U_{\rm ext}(\mathbf{r}), 
 \label{eq:HamiltonianI}\\
 %
  %
&H_{\rm B}=\int d^{d}\mathbf{r}~\Psi ^{\dagger}(\mathbf{r})  \left(\frac{\mathbf{P}_{\rm B}^2}{2m_{\rm B}}+V_{\rm ext}(\mathbf{r})\right)\Psi (\mathbf{r})\nonumber\\
&\hspace{0.5cm}=\sum_{\mathbf{k}}\epsilon_{\mathbf{k}}a_{\mathbf{k}}^{\dagger}a_{\mathbf{k}},
 \label{eq:HamiltonianB}\\
  %
&H_{\rm BB}=g_{\rm B}\int d^{d}\mathbf{r}~\Psi ^{\dagger}(\mathbf{r})  \Psi ^{\dagger}(\mathbf{r})\Psi (\mathbf{r})\Psi (\mathbf{r}) \nonumber\\
&\hspace{0.7cm}=\frac{1}{2V}\sum_{\mathbf{k},\mathbf{k{'}},\mathbf{q}}V_{\rm B}(\mathbf{q})a_{\mathbf{k{'}-q}}^{\dagger}a_{\mathbf{k+q}}^{\dagger}a_{\mathbf{k{'}}}a_{\mathbf{k}},
 \label{eq:HamiltonianBB}\\
 %
&H_{\rm IB}=g_{\rm IB}n_{\rm B}=\frac{1}{V}\sum_{\mathbf{k},\mathbf{q}}V_{\rm IB}\rho_{\rm I}(\mathbf{q})a_{\mathbf{k-q}}^{\dagger}a_{\mathbf{k}}.
 \label{eq:HamiltonianIB}
 \end{align}
In the above expressions,  $V_{\rm ext}(\mathbf{r})$ denotes the external potential experienced by the Bosons which are contained in a (box of) volume $V$of the hyperspace.  From now on, we assume a homogenous BEC, that is, $V_{\rm ext}(\mathbf{r})=0$ along the direction of the impurity motion.  For the impurity, the external potential is $U_{\rm ext}(\mathbf{r})$, and  we will study two cases: a free or a parabolically trapped impurity. The bosonic operators $a_{\mathbf{k}}(a_{\mathbf{k}}^{\dagger})$ destroy (create) a boson of mass $m_{\rm B}$ having wave vector  $\mathbf{k}$ and energy $ \epsilon_{\mathbf{k}}=\left( \hbar k \right)^{2} /(2 m_{B})-\mu$, measured from its chemical potential $\mu$. In addition, the quantities $V_{\rm B}$ and $V_{\rm IB}$ represent the Fourier transform $\mathcal{F}_{\mathbf {q}} [.]$ of the impulsive (contact) boson-boson and impurity-boson interactions respectively.  Their explicit expressions are:
\begin{align}
V_{\rm B} (\mathbf{q} )=g_{\rm B}\mathcal{F}_{\mathbf {q}}  [\delta(\mathbf{r}-\mathbf{r}{'})],
 \label{eq:intractionB}
 \end{align}
 \begin{align}
V_{\rm IB} (\mathbf{q} )=g_{\rm IB}\mathcal{F}_{\mathbf {q}}  [\delta(\mathbf{r}-\mathbf{r}{'})].
 \label{eq:intractionIB}
 \end{align}
Here, the respective coupling strengths are $g_{\rm B}$ and  $g_{\rm IB}$. They are mainly determined by the corresponding scattering lengths and densities \cite{Pitaevskii2003,Lewenstein2012} and their explicit expressions will be given later.  We assume that the impurity density is low enough, which allows us to neglect the terms describing the interaction between impurities. The (dimensionless) density of the impurity in the momentum space is given by
\begin{align}
\rho_{\rm I}(\mathbf{q})=\int_{-\infty}^{\infty}d\mathbf{r}{'}~ e^{-i\mathbf{q}.\mathbf{r}{'}}\delta(\mathbf{r}{'}-\mathbf{r}).
 \label{eq:densityI}
 \end{align}
 
Next, for the sake of completeness, we review how to  construct the Fr\"{o}hlich Hamiltonian, which describes the linear interaction between the motional position quadrature of the impurity and the Bogoliubov bosonic modes of BEC. The goal is to show that such linear regime allows us to model the impurity as a quantum Brownian particle which experiences an effective environment formed by the Bogoliubov Bosonic modes of the BEC (as derived in  \cite{2017Lampo}).  

Given that the Hamiltonian of  the Bosonic interaction is not in bilinear form, we linearize it and  replace the creation and annihilation operators by their average values $\sqrt{N_{0}}$. Below a critical temperature, the atoms mainly occupy the ground state forming a BEC, however, we neglect terms proportional to $N_{\mathbf{k}}$  ($\mathbf{k}\neq0$) i.e. the number of particles out of the ground state. In order to diagonalise the bath modes, we further apply the following Bogoliubov transformation 
\begin{align}
a_{\mathbf{k}}=u_{\mathbf{k}}b_{\mathbf{k}}-v_{\mathbf{k}}b_{\mathbf{-k}}^{\dagger}~, \quad a_{\mathbf{-k}}=u_{\mathbf{k}}b_{\mathbf{-k}}-v_{\mathbf{k}}b_{\mathbf{k}}^{\dagger}.
 \label{eq:BogliubovTran}
 \end{align}
The transformation coefficients are
\begin{align}
u_{\mathbf{k}}^2=\frac{1}{2}\left(\frac{\epsilon_{\mathbf {k}}+n_{0}V_{\rm B}}{E_{\mathbf{k}}}+1\right),
 \label{eq:BogliubovCoef}
 \end{align}
\begin{align}
v_{\mathbf{k}}^2=\frac{1}{2}\left(\frac{\epsilon_{\mathbf {k}}+n_{0}V_{\rm B}}{E_{\mathbf{k}}}-1\right),
 \label{eq:BogliubovCoef}
 \end{align}
 where $n_{0}$ is the (constant) density of particles in the ground state of the homogeneous gas and the Bogoliubov energy spectrum is given by
 \begin{align}
E_{\mathbf{k}}=\hbar\omega_{\mathbf{k}}=\hbar c\left|\mathbf{k}\right|\sqrt{1+\frac{1}{2}\left(\xi \mathbf{k}\right)^2},
 \label{eq:BogliubovSpec}
 \end{align}
with
\begin{align}
\xi=\frac{\hbar}{\sqrt{2g_{\rm B}m_{\rm B}n_{0}}}~, \quad c=\frac{\hbar}{\sqrt{2}m_{\rm B}\xi},
 \label{}
 \end{align}
 representing  the coherence length and the speed of sound, respectively. The effective bath Hamiltonian under such transformation reads \cite{Pitaevskii2003}
 \begin{align}
H_{\rm B}+H_{\rm BB}=\sum_{\mathbf{k}\neq 0}E_{\mathbf{k}} b_{\mathbf{k}}^{\dagger}b_{\mathbf{k}}.
 \label{}
 \end{align}
 Here, we have neglected the non-operator terms which simply shift the energy level of the atoms in BEC. We approximate the bosons-impurity interaction in a similar way.  Further, we only keep the terms proportional to $\sqrt{N_{0}}$, where the macroscopic occupation of the condensate holds as expressed by the condition $ N_{i\neq0}\ll N_{0}$ . By discarding the terms which might cause non-physical instabilities and those bilinear in $\sqrt{N_{0}}$, we obtain the Hamiltonian 
 \begin{align}
H_{\rm IB}=n_{0}V_{\rm IB}+\sqrt{\frac{n_{0}}{V}}\sum_{\mathbf{k}\neq 0}\rho(\mathbf{k})V_{\rm IB}(a_{\mathbf{k}}+ a_{\mathbf{k}}^{\dagger} ).
 \label{eq:HamiltonianIBSimplified}
 \end{align}
 The first term represents simply the constant mean field energy and provides the shift of the energy of polaron.  For the purposes here, it can be neglected. By further invoking the transformation from  Eq. (\ref{eq:BogliubovTran}) into Eq. (\ref{eq:HamiltonianIBSimplified}), one gets 
  \begin{align}
  H_{\rm IB}&=\sqrt{\frac{n_{0}}{V}}\sum_{\mathbf{k}\neq 0}\rho(\mathbf{k})V_{\rm IB} \left(u_{\mathbf{k}}-v_{\mathbf{k}}\right)(b_{\mathbf{k}}+ b_{\mathbf{-k}}^{\dagger} )\nonumber\\
&=\sqrt{\frac{n_{0}}{V}}\sum_{\mathbf{k}\neq 0}\rho(\mathbf{k})V_{\rm IB} \sqrt{\frac{\epsilon_{\mathbf{k}}}{E_{\mathbf{k}}}}(b_{\mathbf{k}}+ b_{\mathbf{-k}}^{\dagger} ),
 \label{eq:HamiltonianIBBoglibovTran}
 \end{align}
where once again we have discarded the non-operator terms. Since the density is dependent on the position of the impurity, we insert its expression into Eq. \eqref{eq:HamiltonianIBBoglibovTran}, which results in the interaction between impurity position and bath variables given by
\begin{align}
  H_{\rm IB}=\sum_{\mathbf{k}\neq 0}V_{\mathbf{k}} e^{i \mathbf{k}\cdot\mathbf{r}}(b_{\mathbf{k}}+ b_{\mathbf{-k}}^{\dagger} ).
 \label{eq:HamiltonianFrohlichInt1}
 \end{align}
Importantly,  $V_{\mathbf{k}}$  contains the impurity-Boson coupling coefficient, and takes the form
\begin{align}
 V_{\mathbf{k}}=g_{\rm IB}\sqrt{\frac{n_{0}}{V}}\left[\frac{(\xi k)^{2}}{(\xi k)^{2}+2}\right]^{\frac{1}{4}}.
 \label{eq:CouplingCoeff}
 \end{align}
The interaction in the Eq. (\ref {eq:HamiltonianFrohlichInt1}) is the interaction part of the  Fr\"{o}hlich Hamiltonian. Under the assumption that one  restricts the calculation  to the limit $\mathbf{k}\cdot \mathbf{r}\ll1$, the interaction reads
 \begin{align}
  H_{\rm IB}=\sum_{\mathbf{k}\neq 0}V_{\mathbf{k}} \left(1 + i \mathbf{k}\cdot\mathbf{r}\right) (b_{\mathbf{k}}+ b_{\mathbf{-k}}^{\dagger} ).
 \label{eq:HamiltonianFrohlichInt2}
 \end{align}
We further simplify it by redefining the Bogoliubov modes operator $ b_{\mathbf{k}}\rightarrow b_{\mathbf{k}}-v_{\mathbf{k}}/E_{k}\mathbf{1}$, to absorb terms proportional to identity operator. After all these simplifications,  the final form of the Hamiltonian of the impurity  in a BEC reads
\begin{align}
  H=H_{\rm I}+\sum_{\mathbf{k}\neq 0}E_{\mathbf{k}}b_{\mathbf{k}}^{\dagger}b_{\mathbf{k}}+\sum_{\mathbf{k}\neq 0}\hbar \vect{g}_\mathbf{k}\cdot\mathbf{r}~\mathbf{\pi_{\mathbf{k}}},
 \label{eq:HamiltonianFrohlichIntFinal}
 \end{align}
with
\begin{align}
   \vect{g}_\mathbf{k}=\mathbf{k}V_{\mathbf{k}}/\hbar  ~, \quad \mathbf{\pi_{\mathbf{k}}}=i\left(b_{\mathbf{k}}-  b_{\mathbf{k}}^{\dagger}\right).
 \label{eq:CouplingParameterSytemBath}
 \end{align}
The Hamiltonian in Eq. (\ref{eq:HamiltonianFrohlichIntFinal}) describes a linear interaction between the impurity center of mass motion and a bath of the Bogoliubov modes of  a BEC. It thus has a form of the QBM Hamiltonian, in which the impurity plays the role of a Brownian particle while the modes of BEC act as an effective Bosonic environment as represented by its (dimensionless) momenta $\mathbf{\pi_{\mathbf{k}}}$. 

\section{\lowercase{d}-Dimensional Spectral Density$-$} \label{sec:SD}
The Hamiltonian derived in the previous section allows us to study the quantum dynamics of an impurity, taking advantage of the analogy with the QBM model.  To characterise the bath, we write its self-correlation function as 
\begin{align}
 \underline{\underline{\mathcal{C}}}(\tau)=\sum_{\mathbf{k}\neq 0}\hbar \underline{\underline{g_{\mathbf {k}}}}\left\langle\mathbf{\pi_{\mathbf{k}}}(\tau)\mathbf{\pi_{\mathbf{k}}}(0)\right\rangle.
 \label{eq:SelfCorrelationBath1}
 \end{align}
Here $\underline{\underline{g_{\mathbf {k}}}}=\underline{g_{\mathbf {k}}}~\underline{g_{\mathbf {k}}}^{T}$ is the coupling tensor.  The environment is made of bosons whose state at finite temperature $T$ follows the Bose-Einstein statistics. Therefore, the mean number of bosons in each of the modes reads  
 \begin{align}
 \langle b_{\mathbf{k}}^{\dagger}b_{\mathbf{k}}\rangle=\frac{1}{\exp(\hbar \omega_{\mathbf{k}}/k_{\rm B}T)-1}.
 \label{eq:CorrelationAnihlationCreation}
 \end{align}
In order to calculate the correlation, we invoke the expression for the dimensionless momenta and make use of  Eq. (\ref{eq:CorrelationAnihlationCreation}) and  Eq. (\ref{eq:SelfCorrelationBath1}) which results in
 \begin{align}
 \underline{\underline{\mathcal{C}}}(\tau)&=\sum_{\mathbf{k}\neq 0}\hbar \underline{\underline{g_{\mathbf {k}}}}\left[\coth\left(\frac{\hbar \omega_{\mathbf{k}}}{2k_{\rm B}T}\right)\cos (\omega_{\mathbf{k}}\tau)-i\sin(\omega_{\mathbf{k}}\tau)\right]\nonumber\\
  &\equiv  \underline{\underline{\nu}}(\tau)-i \underline{\underline{\lambda}}(\tau),
 \label{}
 \end{align}
where the real and imaginary part of the self-correlation function are given by
 \begin{align}
 \underline{\underline{\nu}}(\tau)=\int_{0}^{\infty}  \underline{\underline{J}}(\omega) \coth\left(\frac{\hbar \omega}{2k_{\rm B}T}\right)\cos (\omega\tau) d\omega,
 \label{eq:NoiseKernal}
 \end{align}
\begin{align}
  \underline{\underline{\lambda}}(\tau)=\int_{0}^{\infty} \underline{\underline{J}}(\omega) \sin (\omega\tau) d\omega=-m_{\rm I}\underline{\underline{\dot{\Gamma}}}(\tau).
 \label{eq:SelfCorrelationBath2}
 \end{align}
  Moreover, the damping kernel $\Gamma(t) $ can be obtained from
  \begin{align}
\underline{\underline{\Gamma}}(t)=(1/m_{\rm I}) \int_{0}^{\infty}d\omega(1/\omega) \underline{\underline{J}}(\omega)\cos(\omega t).
 \label{eq:DampingKernal}
 \end{align}
 In the above expressions we have introduced the spectral density $\underline{\underline{J}}(\omega)$, which fully characterises the effects of the bath on the system. This information is contained in the coupling strengths of the various modes of the bath with the system. The spectral density is defined as
\begin{align}
\underline{\underline{J}}(\omega)=\sum_{\mathbf{k\neq0}}\hbar \underline{\underline{g_{\mathbf {k}}}}\delta(\omega-\omega_{\mathbf {k}}).
 \label{eq:Spectral DensityDiscrete1}
 \end{align}
In the present case, the couplings of the impurity (system) and bosons (bath) interaction can be derived  from  first principles. It is therefore possible to obtain the exact expression for the spectral density. This scenario is in contrast to various complicated system-bath interactions, where it is hard to get an exact form of the spectral densities, such as in bulk mechanical structure akin to opto-mechanical setup \cite{RevModPhys.86.1391, PhysRevA.94.063830, Grblacher2015}.  While the case of spectral density in $1d$ has been studied in \citep{2017Lampo}, here we derive it systematically in $d=\{1,2,3\}$ dimensions of the quasi momentum space. In appendix \ref{App:AA}, we derive the expression for the spectral density tensor, which is given by
\begin{align}
\underline{\underline{J_{\rm d}}}(\omega)= d^{-1}\pr{\mathcal{J}_{\rm d}(\omega)} \underline{\underline{I} }~_{\rm d\times d},
\label{eq:JTensorMainText}
 \end{align}
where $ \underline{\underline{I} }~_{\rm d\times d}$  is the identity matrix and the scalar function $ \mathcal{J}_{\rm d}(\omega)$ in d dimensions is given by
\begin{align}
&\mathcal{J}_{\rm d}(\omega)=\pc{\frac{S_{\rm d} \pc{\sqrt{2}}^{d}(\eta_{\rm d})^{2}(\Lambda_{\rm d})^{d+2}}{(2\pi)^d}}\times\nonumber\\
&\pc{\frac{\left[\left(\frac{m_{\rm B}}{\pr{g_{\rm B, d}}^{\pr{\frac{d}{d+2}}}n_{0,\rm d}}\right)\left(\sqrt{\frac{\omega ^2}{(\Lambda_{\rm d}) ^2}+1}-1\right)\right]^{(\frac{d+2}{2})}}{\left(\sqrt{\frac{\omega ^2}{(\Lambda_{\rm d})^2}+1}\right) }}.
\label{}
\end{align}
For $ d=1, ~2$ and  $3$ we have $S_{1}=2,~ S_{2}=2\pi$ and $S_{3}=4\pi$ respectively. Moreover, we have defined the $d-$dependent characteristic frequency $ \Lambda_{\rm d}=(g_{\rm B,d}n_{0,{\rm d}})/\hbar$ because the  boson-boson coupling and the density  differ in various dimensions.  We also write the impurity-boson coupling  in the units of the boson-boson coupling as $\eta_{\rm d}=(g_{\rm IB,d}/g_{\rm B,d})$. Such characterization allows us to study the long-time dynamics of the impurity in the following sense:  one can identify two opposite limits in the above expression i.e. $\omega\ll\Lambda_{\rm d}$ and $\omega\gg\Lambda_{\rm d}$ in which $\Lambda_{\rm d}$ appears naturally as the characteristic cut-off frequency which distinguishes between the low and the high frequencies of the bath. The low-frequency behaviour is attributed to the linear part of the Bogoliubov spectrum \cite{2017Lampo}.  From the Tauberian theorem \cite{nixon1965handbook}, one can obtain the long-time behaviour of a function which is determined by the low frequency response of its Laplace transform. The above low-frequency choice is therefore a natural way of studying the dynamics perturbed by the bath that acts beyond the very short transient regime. Note that, for $ d=1$ the above expression reduces to the one dimensional spectral density used in \citep{2017Lampo}. To the lowest order of $ \omega/\Lambda_{\rm d}$, the expression for the  spectral density with the low frequency response of the bath is given by
\begin{align}
\mathcal{J}_{\rm d}(\omega)\simeq\frac{S_{\rm d}\pc{ \eta_{\rm d}}^{2}}{2(2\pi)^d}\left(\frac{m_{\rm B}}{\pr{g_{\rm B, d}}^{\pr{\frac{d}{d+2}}}n_{0,\rm d}}\right)^{(\frac{d+2}{2})}\times\omega^{d+2}.
\label{eq:Spectral DensityFuction}
\end{align}
This expression gives the scaling of the frequency for the spectral density function in all dimensions. We point out that due to the spherical symmetry of the bath, the spectral density tensor is a diagonal matrix. As a consequence, the noise and damping kernels given by Eq. (\ref{eq:NoiseKernal}) and Eq. (\ref{eq:DampingKernal}) are also diagonal. We now give further details about the other parameters involved. In dimension $d$, the coupling constant $g_{\rm B,{\rm d}}$ and boson density $n_{0,{\rm d}}$ have the form
\begin{align}
g_{\rm B,{\rm d}}=\frac{S_{\rm d}\hbar^{2}a_{3}}{m_{\rm B}\pc{\sqrt{\hbar/m_{\rm B}\omega_{\rm d}}}^{3-d}},~ \quad n_{0,{\rm d}}=\pc{n_{0,{\rm 1}}}^{d},
\label{eq:BosonCouplingConstant&density}
\end{align}
their units being  $ { \rm J\cdot m}^{d}$ and  $ {\rm m}^{-d}$ respectively. Here we have written these expressions in terms of the three-dimensional scattering length $a_{3}$  and one-dimensional density $n_{0,1}$. We have further assumed  transverse confinement of the boson gas with a harmonic trap having a Gaussian ground state \cite{PhysRevA.85.063626}, which makes the cases $d<3$ to be the quasi one- and two-dimensional.  We emphasize that the parabolic potential is introduced only in the direction transverse to the direction under investigation. The dynamics we study is thus still confined to a box potential, making the homogeneity of Boson gas to be a valid approximation. Moreover, the zero point fluctuations of the condensate are characterised  by the trapped frequencies  $\omega_{\rm d}=\px{\omega_{1}=\omega_{\perp},\omega_{2}=\omega_{z}, \omega_{3}=0}$. For instance, when we consider one, two- and three-dimensional condensate to be confined in the $x$ direction, in the  $x-y$ plane or in the volume $x-y-z$ respectively, the explicit form of the potential may be given by
\begin{align}
V_{\rm ext}(\mathbf{r})=\begin{cases}
   (1/2)m_{\rm B}\omega_{\perp}^{2}\left(y^{2}+z^{2}\right) ,\quad  \text{for } d=1\\
   (1/2)m_{\rm B}\omega_{z}^{2}\left(z^{2}\right),\quad  \quad  \text{for }~ d=2\\
0~ ,\quad \quad\quad \quad \quad \quad \quad  \text{for }~ d=3.\\
\end{cases}
 \label{}
 \end{align}
Note that for $d=3$ there is no parabolic confinement and therefore the expressions are independent of the trapping frequency. It is then possible to define a characteristic time $\tau_{\rm d}$ which is raised to the power $d$ in the expression for the spectral function, which is given by%
\begin{align}
&\mathcal{J}_{\rm d}(\omega)=m_{\rm I}\pc{\tau_{\rm d}}^{d}\omega^{d+2}\quad \text{where} \nonumber\\ 
&\pc{\tau_{\rm d}}^{d}\equiv\frac{S_{\rm d} \pc{\eta_{\rm d}}^{2}}{2(2\pi)^dm_{\rm I}}\left(\frac{m_{\rm B}}{\pr{g_{\rm B, d}}^{\pr{\frac{d}{d+2}}}n_{0,\rm d}}\right)^{(\frac{d+2}{2})}.
\label{eq:Spectral DensityDiscrete}
\end{align}
It is evident that the spectral density has a  super-ohmic dependence on the frequency in all dimensions.   It is  therefore expected that the bosonic bath would induce a non-Markovian dynamics of the impurity. Moreover, it can be shown that the increasing nature of the spectral density makes certain quantities, such as momentum dispersion, diverge. It is therefore customary to define the ultraviolet cut-off $ \mathcal{K} (\omega,\Lambda_{\rm d})$ in order to suppress the contribution of high frequencies. After this, the expression for the spectral density reads
\begin{align}
\mathcal{J}_{\rm d}(\omega)=m_{\rm I}\pc{\tau_{\rm d}}^{d}\omega^{d+2}\mathcal{K} (\omega,\Lambda_{\rm d}).
\label{eq:Spectral DensityFuctionApprox}
\end{align}
In the following we study the impurity dynamics, varying the dimension and d-dependent cut-off function in the expression of the spectral density.

\section{Dynamics and Control $-$}\label{sec:dynamics}
To study the quantum dynamics of the impurity atom, we write down the equations of motion in the Heisenberg picture. The impurity, which is immersed in a bath of dimension $d$,  is further trapped by a harmonic potential. In dimensions $1$, $2$ and $3$, the potential is  $U_{\rm ext}(x)=(1/2)m_{\rm I}\Omega^{2}\pc{x^{2}}$, $U_{\rm ext}(x,y)=(1/2)m_{\rm I}\Omega^{2}\pc{x^{2}+y^{2}}$ and $U_{\rm ext}(x,y,z)=(1/2)m_{\rm I}\Omega^{2}\left(x^{2}+y^{2}+z^{2}\right)$ respectively. Here, we have assumed equal trapping frequency in all the directions available to the impurity dynamics. The free QBM is therefore characterised by setting  $\Omega=0$ in all the cases. We write the equation of motion in vectorial form as
  \begin{align}
\underline{\dot X}(t)=\frac{i}{\hbar}\left[H,\underline{ X}(t)\right]=\frac{\underline{\dot P}(t)}{m_{\rm I}},
 \label{eq:EOMPositionImpurity}
 \end{align}
  \begin{align}
\underline{\dot P}(t)=\frac{i}{\hbar}\left[H,\underline{P}(t)\right]=-m_{\rm I}\Omega^{2}\underline{\dot X}(t)-\hbar \sum_{k}\underline{g_{k}}\pi_{k}(t),
 \label{eq:EOMMomentumImpurity}
 \end{align}
  \begin{align}
\dot b_{k}(t)=\frac{i}{\hbar}\left[H,b_{k}(t)\right]=-i\omega_{k}b_{k}(t)-\underline{g_{k}}^{T} \underline{X}(t),
 \label{eq:EOMBogoliubovOperator}
 \end{align}
 \begin{align}
\dot b_{k}^{\dagger}(t)=\frac{i}{\hbar}\left[H,b_{k}^{\dagger}(t)\right]=-i\omega_{k}b_{k}^{\dagger}(t)-\underline{g_{k}}^{T} \underline{X}(t).
 \label{eq:EOMBogoliubovOperatorTC}
 \end{align}
Here $H$ represents the  Hamiltonian of the system given by Eq. (\ref{eq:HamiltonianFrohlichIntFinal}). In general, the dimension of the vectors in the above equations is $d$, the dimension of the bath.  
In appendix \ref{App:BB}, we  combined these equations to obtain an equation of motion  for the impurity position vector:
\begin{align}
&\underline{\ddot X}(t)+\Omega^{2}\underline{X}(t)+\partial_{t}\int_{0}^{t}\underline{\underline{\Gamma}}(t-s)\underline{X}(s)ds\nonumber\\
&=(1/m_{\rm I})\underline{B}(t).
 \label{eq:EOMPositionImpuritydynamicalmap}
 \end{align} 
Here, the quantum Brownian stochastic force  $\underline{B}(t)$ stands for
\begin{align}
\underline{B}(t)=\sum_{k}i\hbar \underline{g_{k}}(b_{k}^{\dagger}e^{i\omega_{k}t}-b_{k}e^{-i\omega_{k}t}).
 \label{eq:BathStochasticForce}
 \end{align}
In any given dimension, the diagonal damping kernel when equal weighting for all the directions is taken, will suffice to study the motion along any one of the coordinate axis. However for different dimensions, the spectral density will bring different level of {\it super-ohmicity}, as stated by Eq. (\ref{eq:Spectral DensityFuctionApprox}). Therefore, the form of the noise and damping kernels will also differ according to the dimension involved.   As a result, the impurity motion is different for different dimensions, despite being studied along one particular coordinate axis.  One can then aim to study such a motion by constructing a unit vector $\pc{1,0,0}$ (i.e. along the x direction) and taking dot product with Eq. (\ref{eq:EOMPositionImpuritydynamicalmap}). This results in
 \begin{align}
\ddot x(t)+\Omega^{2} x(t)+\partial_{t}\int_{0}^{t}\Gamma_{\rm d}^{xx}(t-s)x(s)ds
=\pc{\frac{1}{m_{\rm I}}}B^{x}(t).
 \label{eq:EOMPositionImpurityXdirection}
 \end{align} 
Note that the tensor components satisfy $ \Gamma_{\rm d}^{xy}= \Gamma_{\rm d}^{xz}=0$. Additionally, from the structure of the integral term in the Eq. \eqref{eq:EOMPositionImpurityXdirection}, it is obvious that damping kernel is non-local in time. This implies that the dynamics of the impurity  depends on its history. Therefore, in general the impurity motion displays memory effects. Only in the case of Ohmic spectral density (linear function of $\omega$) the memory damping kernel reduces to a Dirac delta function and describes the time-local dynamics of the standard damped harmonic oscillator. The time local behaviour is violated in similar experimental configuration \cite{Grblacher2015} and surge of non-Markovianity is addressed elsewhere \cite{minoguchi2019environment, reitz2019molecule, PhysRevA.93.063853}.
The formal solution to Langevin-like Eq. (\ref{eq:EOMPositionImpurityXdirection}) takes the form
\begin{align}
x(t)=&G_{\rm 1,d}(t)x(0)+G_{\rm 2,d}(t)\dot x(0)+(1/m_{\rm I})\int_{0}^{t}ds\nonumber\\
&\times G_{\rm 2,d}(t-s)B^{x}(s),
 \label{eq:SolutionEOM}
 \end{align}
where the Green's functions $G_{\rm 1,d}$ and $G_{\rm 2,d}$ are defined in terms of their Laplace transforms 
\begin{align}
\mathcal{L}_{\mathcal{S},{\rm d}}\left[G_{\rm 1,d}(t)\right]=\frac{\mathcal{S}}{\mathcal{S}^{2}+\Omega^{2}+\mathcal{S}\mathcal{L}_{\mathcal{S}, {\rm d}}\left[\Gamma_{\rm d}^{xx}(t)\right]},
 \label{eq:LaplaceG1}
 \end{align}
 \begin{align}
\mathcal{L}_{\mathcal{S},{\rm d}}\left[G_{\rm 2,d}(t)\right]=\frac{1}{\mathcal{S}^{2}+\Omega^{2}+\mathcal{S}\mathcal{L}_{\mathcal{S}, {\rm d}}\left[\Gamma_{\rm d}^{xx}(t)\right]}.
 \label{eq:LaplaceG2}
 \end{align}
 Moreover,  they satisfy the following initial conditions
 \begin{align}
G_{\rm 1,d}(0)=1, ~~~ \dot{G}_{\rm 1,d}(0)=0,
 \label{eq:InitialconditionG1}
 \end{align}
  \begin{align}
G_{\rm 2,d}(0)=0, ~~~ \dot{G}_{\rm 2,d}(0)=1.
\label{eq:InitialconditionG2}
 \end{align}
 
 \begin{figure}[t!]
\includegraphics[width=0.9\linewidth]{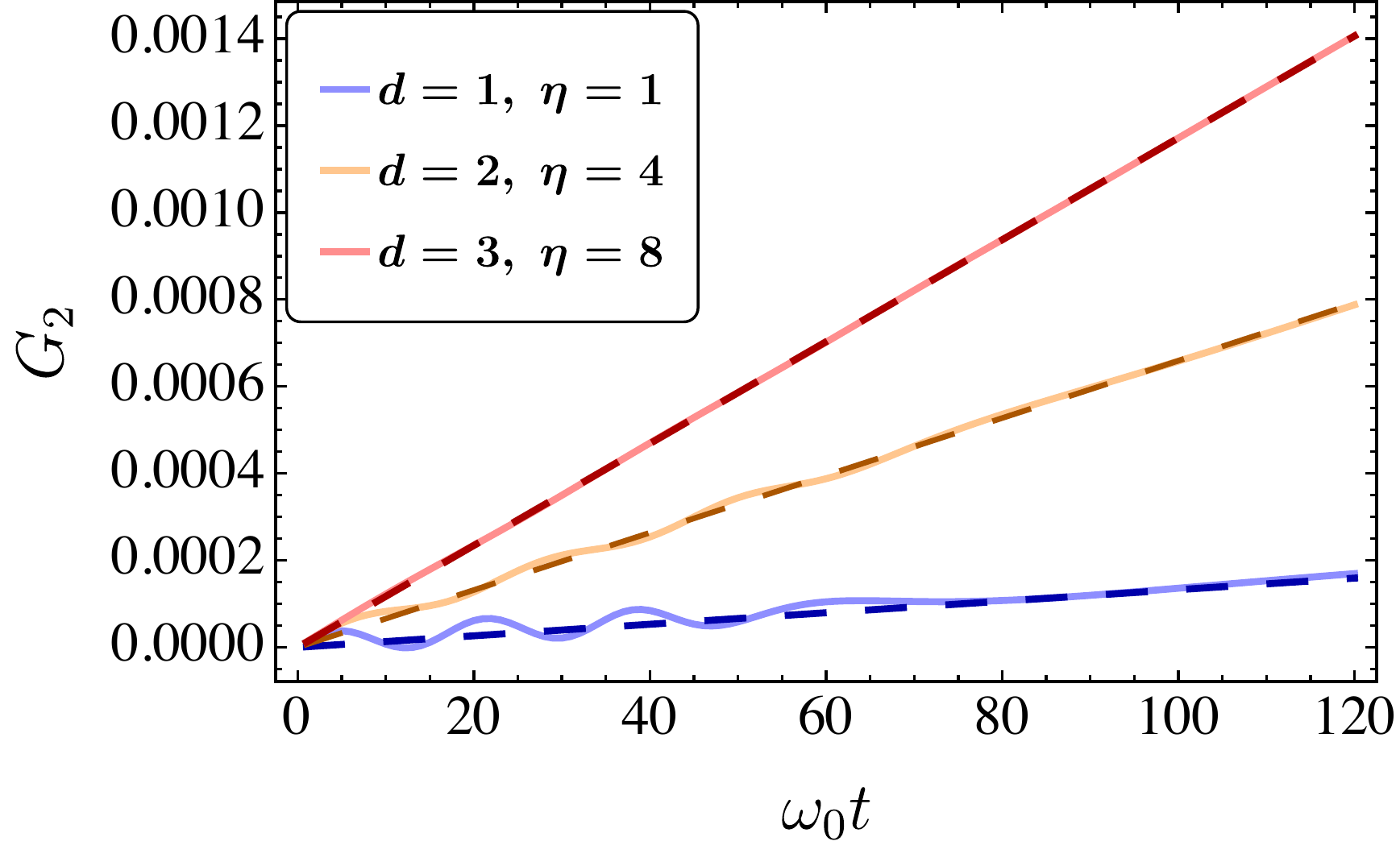}
\caption{(Color online)  Dynamics of the propagator $G_{2}(t)$ in the untrapped case for various dimensions. The dark dashed lines show the asymptotic behaviour given by Eq. (\ref{eq:G1G2Asymtotic}), while the light solid lines represent the solution obtained by employing the Zakian method. In the long time limit, both  solutions match for all dimensions. The results refer to an impurity ${\rm K}$ with mass $m_{\rm I} = 6.4924249\times10^{-26} \rm{kg}$, immersed in a gas of $\rm Rb$  with mass $m_{\rm B} = 1.4192261\times10^{-25} \rm{kg}$. The one-dimensional boson density is $n_{0,1}=7(\mu{ \rm m})^{-1}$. We fix the three-dimensional scattering length $a_{3}=100 ~a_{0}$, where $a_{0}$ is the Bohr radius. Here, the time axis is scaled with the one-dimensional characteristic frequency $ \omega_{0}\equiv(\hbar n_{0,1}^{2}/m_{\rm I})$. }
\label{fig:fig1}
\end{figure}

From now on, the dynamics in the case when high frequencies are suppressed, will be analyzed by introducing a sharp cutoff. The latter is given by $ \mathcal{K}=\Theta (\Lambda_{\rm d}-\omega)$, where $\Theta$ is the Heaviside step function. We further use Eq.~\eqref{eq:DampingKernal} to compute the damping kernel in any dimension $d$
 \begin{align}
\mathcal{L}_{\mathcal{S},{\rm d}}\left[\Gamma_{\rm d}^{xx}(t)\right]=\frac{\pc{\Lambda_{\rm d}} ^{d+2} \pc{\tau_{\rm d}} ^d \,_2F_1\left(1,\frac{d+2}{2};\frac{d+4}{2};-\frac{\pc{\Lambda_{\rm d}} ^2}{\mathcal{S}^2}\right)}{d(d+2)\mathcal{S}},
 \label{eq:LaplaceDamping}
 \end{align}
 with $ \,_2F_1\left[.\right]$ denoting the hypergeometric function. 
 
 \subsection{Untrapped Case}
 Let us first study the free QBM, that is the untrapped case  $\Omega=0$. The quantities of interest in this case are the mean squared displacement ${\rm MSD_{d}}(t)$ (see definition below, Eq. \eqref{eq:MSD}) and the average kinetic energy $E_{\rm d}(t)$ of the impurity. The motion is fully characterised by the functions $G_{1}(t)$ and $G_{2}(t)$ which are the inverse Laplace transform of Eq. (\ref{eq:LaplaceG1}) and  Eq. (\ref{eq:LaplaceG2}) respectively. Exact analytical expressions for these functions are hard to obtain.  Note however, that the Laplace transform of both of these functions are expressed in terms of the Laplace transform of the damping kernel. In the regime of interest $|\mathcal{S}| \ll\Lambda_{\rm d}$, which characterises the low frequency response, we have approximately
  \begin{align}
\mathcal{L}_{\mathcal{S},{\rm d}}\left[\Gamma^{xx}_{\rm d}(t)\right]=d^{-2}(\Lambda_{\rm d}\tau_{\rm d})^d \mathcal{S}.
 \label{eq:}
 \end{align}
 We therefore obtain the asymptotic expressions for the Laplace transforms of the position and momentum propagators
\begin{align}
\mathcal{L}_{\mathcal{S}, \rm d}\left[G_{1,\rm d}(t)\right]=\frac{1}{\alpha_{\rm d} \mathcal{S}} \quad \text{and}\quad \mathcal{L}_{\mathcal{S}}\left[G_{2,\rm d}(t)\right]= \frac{1}{\alpha_{\rm d}\mathcal{S}^{2}},
 \end{align}
 where $\alpha_{\rm d}= 1+d^{-2}(\Lambda_{\rm d}\tau_{\rm d})^d $. Their time domain representations read
 \begin{align}
G_{1, \rm d}(t)=1/\alpha_{\rm d}~ , \quad G_{2,\rm d}(t)=t/(\alpha_{\rm d}).
\label{eq:G1G2Asymtotic}
 \end{align} 
We note that such expressions do not satisfy the boundary conditions stated in  Eq. \eqref{eq:InitialconditionG1} and Eq. \eqref{eq:InitialconditionG2}. However, this is justified since the above solution refers to long-time behaviour.  Several algorithms exist for the numerical computation of the inverse Laplace transform of an arbitrary function. Here we employ the  Zakian method  \cite{WANG201580} to compute the inverse Laplace transform of  Eq. (\ref{eq:LaplaceG1}) and  Eq. (\ref{eq:LaplaceG2}). This method approximates the inverse Laplace transform $f(t)$ of a function  $F(\mathcal{S})$ through
\begin{align}
f(t)\simeq\frac{2}{t}\sum_{j=1}^{N}{\Re} \left[k_{j}F(\Xi_{j}/t)\right],
 \end{align} 
with  the values of the complex parameters $k_{j}$ and $\Xi_{j}$ given in Ref. \cite{WANG201580}. In order to check the equivalence between the asymptotic form of $G_{2}(t)$ and its Zakian approximation, we plot both in Fig.~\ref{fig:fig1}. It turns out that they agree in the long time limit. From here on we will be employing them interchangeably according to our computational convenience.  It is evident from Eq.~\eqref{eq:SolutionEOM} that the function $G_{2}(t)$ is responsible for the propagation of the initial velocity of the impurity. From the results shown in Fig.~\ref{fig:fig1}, such a function follows a ballistic profile in any  dimension, i.e. it is a linear function of time. 

 \subsubsection{Mean Square Displacement}
In this section we discuss the mean squared displacement (MSD) of the impurity motion which is a measurable quantity in cold-atom experiments \cite{ Catani2012}. The MSD is defined as 
 \begin{align}
{\rm MSD}_{x, \rm d} (t)=\av{\pr{x(t)-x(0)}^2}.
\label{eq:MSD}
 \end{align} 
The expression of  the MSD for the generalised Heisenberg-Langevin equations is given in the Appendix~\ref{App:CC}.  For the system at hand, we employ the asymptotic expressions of the Green's functions to evaluate the MSD in different dimensions. Its dynamical part is given by
  \begin{align}
{\rm MSD}_{x, \rm d} (t)=&\pc{\frac{t}{\alpha_{\rm d}}}^{2}\langle \dot{x}(0)^{2}\rangle+\frac{1}{2(\alpha_{\rm d} m_{\rm I})^2}\int_{0}^{t}du\int_{0}^{t}dv \nonumber\\
&\times(t-u) (t-v)\langle\{B^{x}(u),B^{x}(v)\}\rangle.
 \end{align}
Additionally, using the diagonal form of the noise tensor given in Eq. \eqref{eq:NoiseKernal}, one can obtain the relation between the  correlation characterised by the positive commutator $\langle\{B^{x}(u),B^{x}(v)\}\rangle$ and the component  $\nu^{xx}(t)$ of the noise kernel (fluctuation-dissipation relation) \cite{2019LampoBook} .  Explicitly,
 \begin{align}
\langle\{B^{x}(u),B^{x}(v)\}\rangle=2\hbar\nu^{xx}(u-v).
\label{symtricBSFCorrelation}
 \end{align}
Substituting the $d$-dimensional spectral function with a sharp cutoff  into the noise component we get
  \begin{align}
&{\rm MSD}_{x, \rm d} (t)=\pc{\frac{t}{\alpha_{\rm d}}}^{2}\langle \dot{x}(0)^{2}\rangle+\frac{\hbar~ d^{-1}\pc{\tau_{\rm d}}^d}{ m_{\rm I}(\alpha_{\rm d} )^2}\int_{0}^{t}du\int_{0}^{t}dv \nonumber\\
&\int_{0}^{\Lambda_{\rm d}}d\omega(t-u) (t-v)\coth{\pc{\frac{\hbar\omega}{2 k_{\rm B}T}}}\cos{\pr{\omega\pc{u-v}}} \omega^{d+2}.
 \end{align}
By performing the two-dimensional integration over the time variables $u$ and $v$, followed by an integration over the variable $\omega$, we evaluate the expression for the low temperature regime, where $\coth\left(\hbar\omega/2 k_{\rm B}T\right)\rightarrow1$ holds.  In the long time limit, the resulting expression for the MSD is dominated by the terms proportional to $t^{2}$ and its explicit expression turns out to be
   \begin{align}
&{\rm MSD}_{x, \rm d}^{\rm LT} (t)=\left[\langle \dot{x}(0)^{2}\rangle+\frac{\hbar\pc{\tau_{\rm d}}^{d}\pc{\Lambda_{\rm d}}^{d+1}}{m_{\rm I}d(d+1)}\right]\left(\frac{t}{\alpha_{\rm d}}\right)^{2}
 \end{align}
In the regime which fulfills the conditions stated above, the MSD is proportional to the square of the time for all dimensions. In the normal diffusion scenario, the MSD  shows a linear dependence on time. If, on the contrary, the MSD is non-linear in time, proportional to $t^ \alpha$ with an exponent  higher than one,  the diffusion is called anomalous and the motion is called superdiffusive. In the present case, superdiffusion is a consequence of the super-ohmic spectral density in every dimension. The coefficient  in the second term is called the superdiffusion coefficient $D_{x, \rm d}$ and can be interpreted as the average of the square of the speed with which the impurity runs away. We thus have
\begin{align}
&D_{x, \rm d}^{\rm LT}=\frac{\hbar\pc{\tau_{\rm d}}^{d}\pc{\Lambda_{\rm d}}^{d+1}}{m_{\rm I}d(d+1)\pc{\alpha_{\rm d}}^{2}}.
 \end{align}
\begin{figure}[t!]
\includegraphics[width=0.9\linewidth]{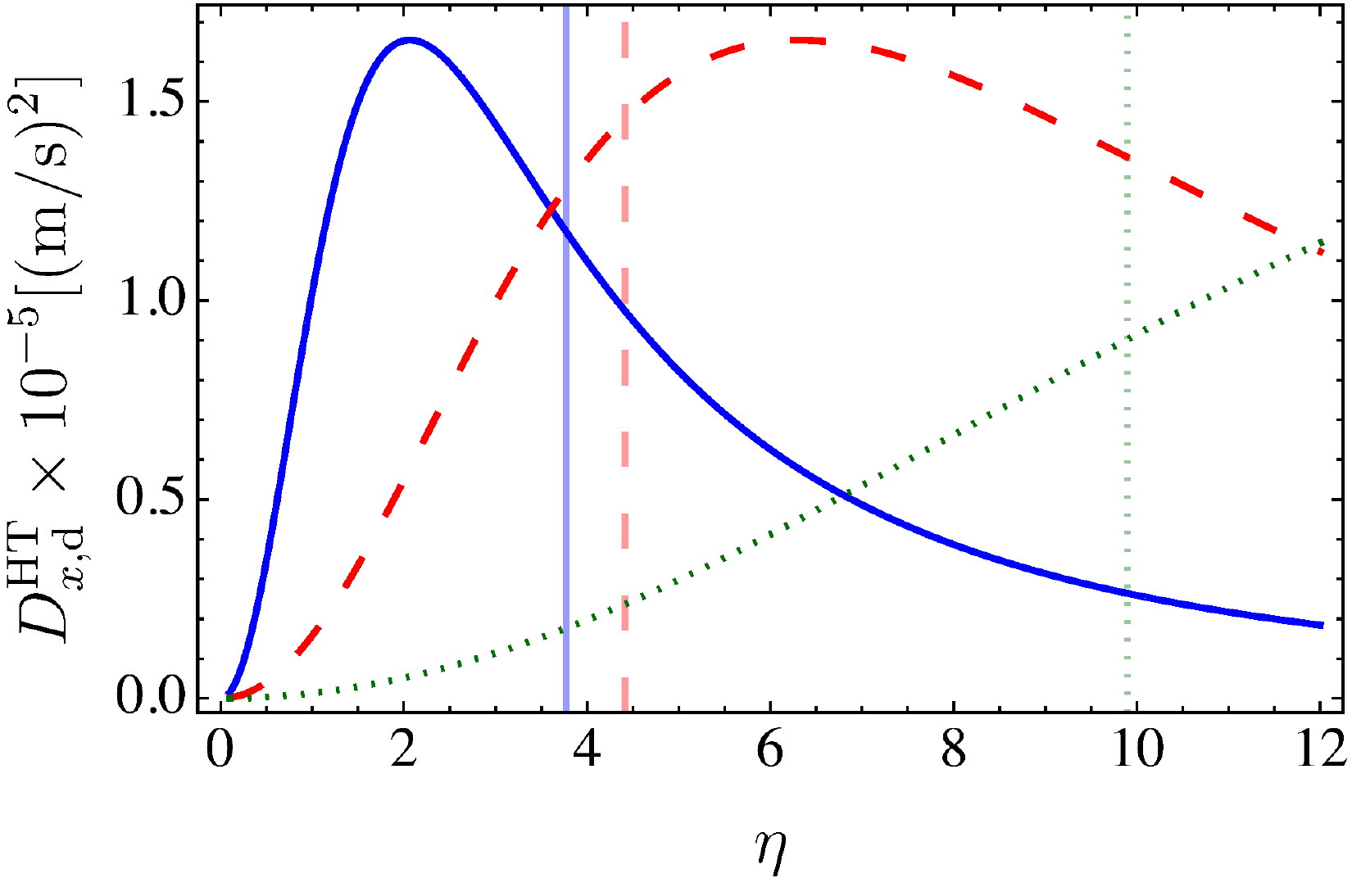}
\caption{(Color online) High temperature super diffusion coefficient in various dimensions as a function of the coupling strength. The solid, dashed and dotted curves represent the $d=1$, $d=2$ and $d=3$ cases respectively. We set temperature $ T=0.15{\mu {\rm K}}$, which fulfills the high temperature condition $k_{\rm B}T\geqslant { \rm Max}\pr{\hbar \Lambda_{\rm d}}$ (see text). The rest of the parameters are the same as in Fig.~\ref{fig:fig1}.  The vertical solid, dashed and dotted lines fix the critical coupling for the Fr\"{o}hlich  Hamiltonian to be valid in one, two and three dimensions, respectively.}
\label{fig:fig2}
\end{figure}
One can perform a similar analysis of the high temperature regime, which is followed by the approximation $\coth\left(\hbar\omega/2 k_{\rm B}T\right)\rightarrow\pc{2k_{\rm B}T/\hbar\omega}$. We remark that the condition $k_{\rm B}T\geqslant { \rm Max}\pr{\hbar \Lambda_{\rm d}}$ implies the high temperature regime in any dimension.   Here ${ \rm Max}[.] $ is the maximum of the cut-off frequencies of different dimensions. This means that all the Bogoliubov modes of the bath will be thermally populated in any of the considered dimensions.   However, while the cut-off frequency in dimension $d$ scales as $\Lambda_{\rm d}\sim \pc{n_{0,1}}^d$, it also depends on the boson coupling constant in the corresponding dimension and therefore on the transverse confinement of the boson gas [cf. Eq. \eqref{eq:BosonCouplingConstant&density}].   Inserting the values of the parameters used in this article, we obtain $ \Lambda_{2}>\Lambda_{3}> \Lambda_{1}$.  The high temperature regime holds as long as $k_{\rm B}T\geqslant\hbar \Lambda_{2}$.  In this regime, the MSD again scales with the square of the time.  The dimension-dependent superdiffusion coefficient  takes the form
\begin{align}
&D_{x, \rm d}^{\rm HT}=\frac{ 2k_{\rm B}T\pc{\tau_{\rm d}}^{d}\pc{\Lambda_{\rm d}}^{d}}{m_{\rm I}d^2\pc{\alpha_{\rm d}}^{2}}.
\end{align}
It is clear from this expression that the superdiffusion coefficient is proportional to the temperature of bath and inversely proportional to the mass of the impurity.
  \begin{figure*}[]
 \includegraphics[width=0.68\columnwidth]{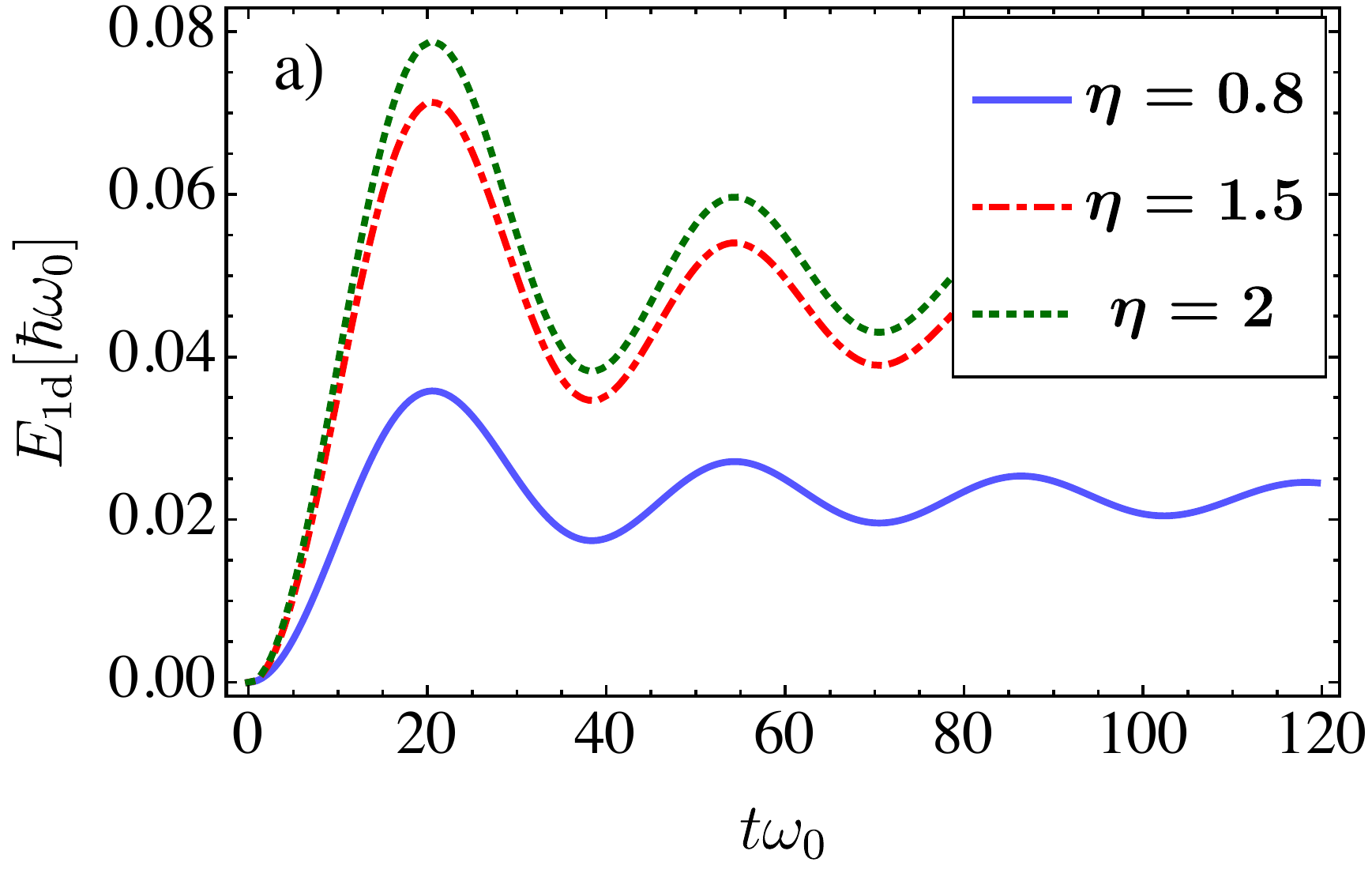}
 \includegraphics[width=0.68\columnwidth]{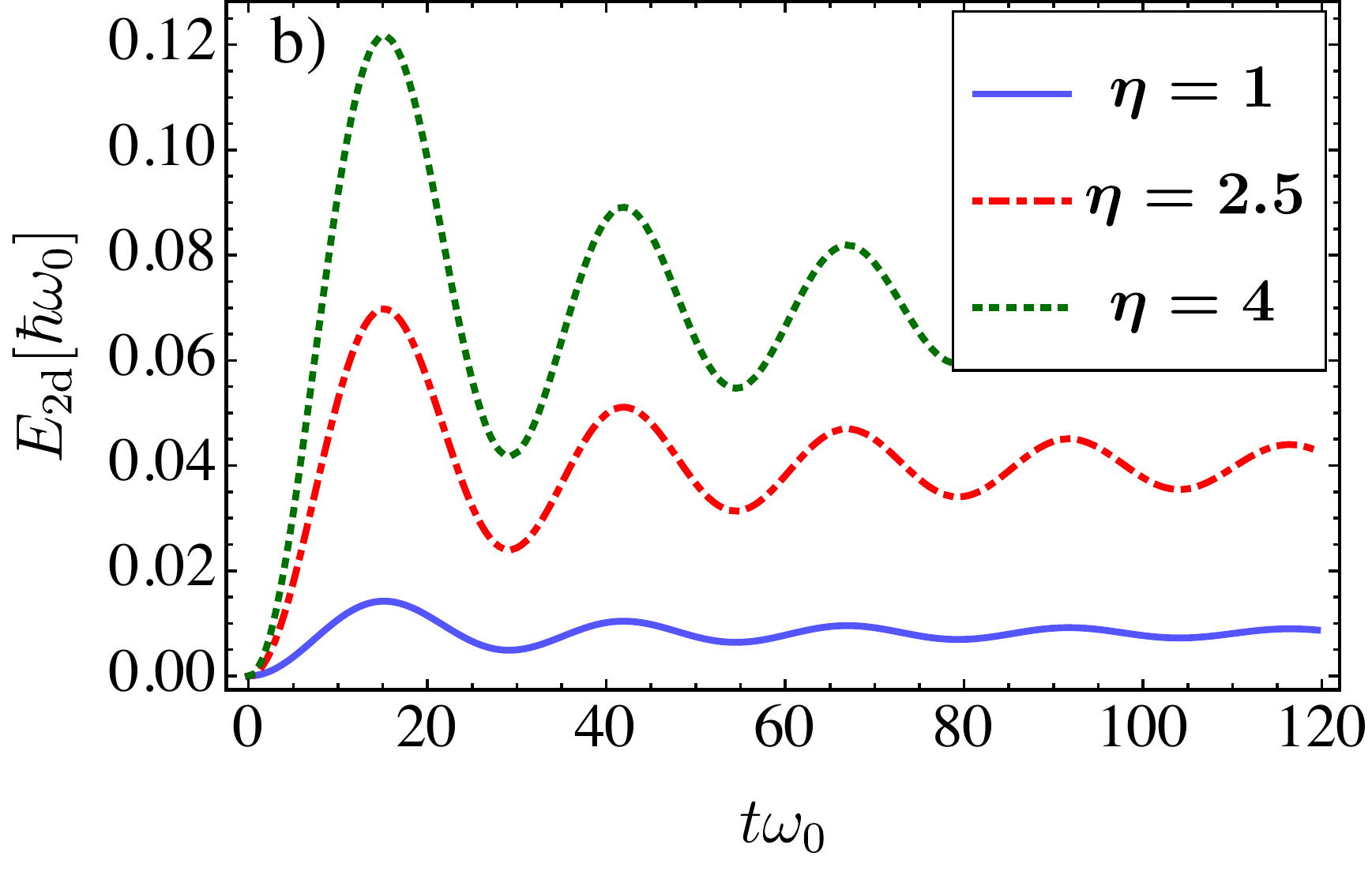}
\includegraphics[width=0.68\columnwidth]{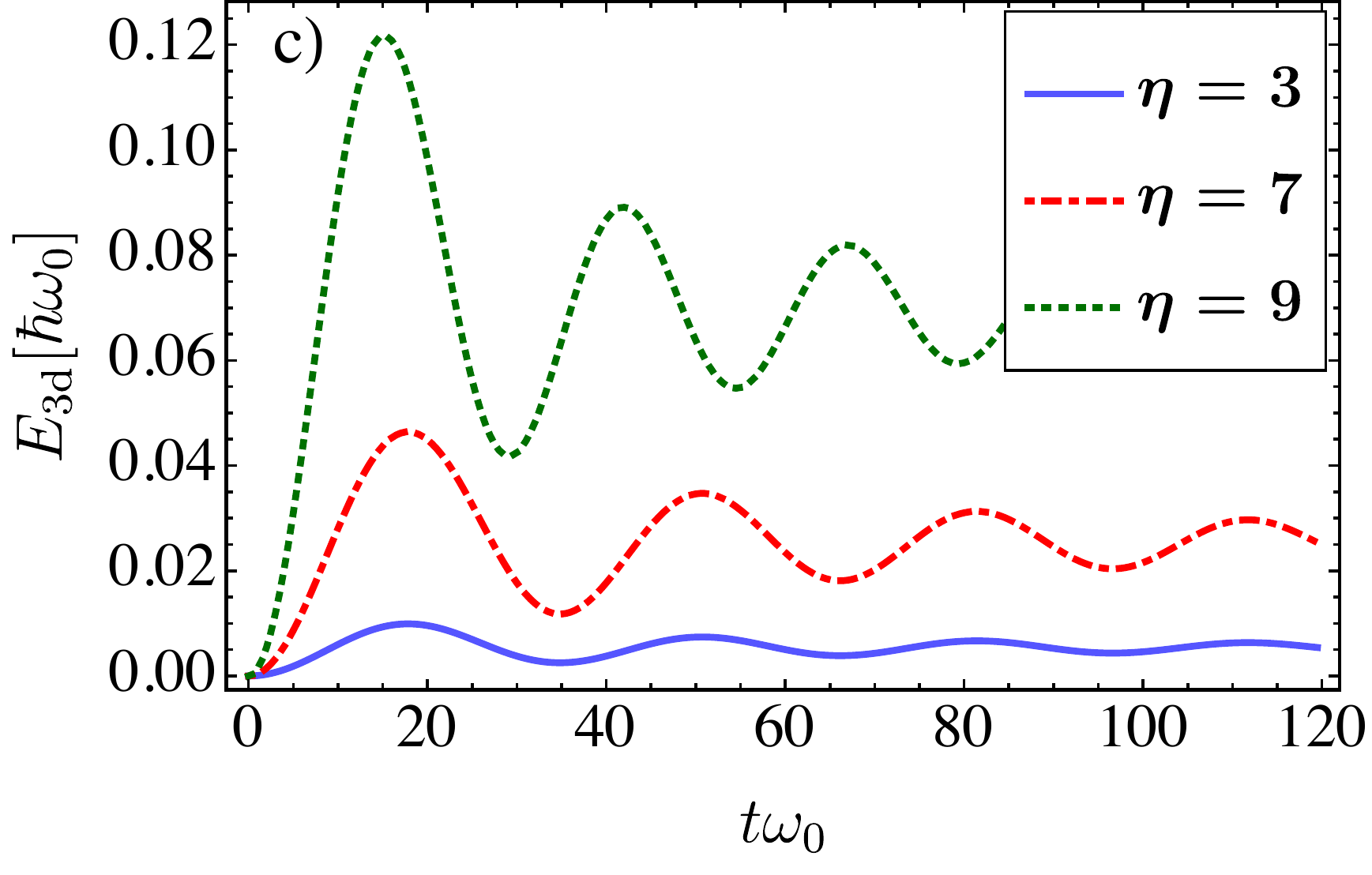}
  \caption{ (a-c): Dynamics of the average energy for 1$d$, 2$d$ and  3$d$, respectively. The coupling strengths are shown in the legend for every case. The rest of the parameters are kept same as before.}
   \label{fig:fig3}
\end{figure*}
One can further write the high temperature superdiffusion coefficient as an explicit function of the coupling parameter:
\begin{align}
D_{x, \rm d}^{\rm HT}=\pr{\frac{ 2k_{\rm B}T}{{m_{\rm I} }}}\pr{ \frac{ \beta_{\rm d}\eta^2}{\pc{\beta_{\rm d}}^{2}d^{-2}\eta^4+2\beta_{\rm d}\eta^2+d^{2}}}.
\end{align}
Here, we have defined the quantities
\begin{align}
\beta_{\rm d}\equiv\pc{\Lambda_{\rm d}\tau_{\rm d, s}}^d\quad \text{where}\quad \pc{\tau_{\rm d, s}}^{d}\equiv \eta^{-2}\pc{\tau_{\rm d}}^{d}.
\end{align}
In dimension $d$, the maximum of this function occurs at
\begin{align}
\eta_{\rm max, d}=\frac{d}{\sqrt{\pc{\Lambda_{\rm d}\tau_{\rm d, s}}^{d}}}.
\end{align}
and has the same maximal value in all dimensions:
\begin{align}
D_{x, \rm d}^{\rm HT, \rm max}=\pc{\frac{1}{m_{\rm I}}}\pc{\frac{k_{\rm B }T}{2}}.
\end{align}
In Fig. \ref{fig:fig2}, we plot the high temperature superdiffusion coefficient for a range of coupling strengths, covering the allowed critical coupling  in every dimension (see appendix \ref{App:DD} for the validity of  the Fr\"{o}hlich Hamiltonian). In all dimensions, for sufficiently weak coupling strengths, one observes a corresponding increase of the bath-induced momentum diffusions of the impurity as the coupling strength increases. However, above $\eta_{\rm max, d}$, the super diffusion coefficient is reduced as the coupling grows. Such a damped regime occurs only in the one-dimensional case within the Fr\"{o}hlich regime, where the impurity may exert both underdamped and overdamped motion.  (In higher dimensions the over damped regime occurs past the vertical line that signals the value of the coupling, critical for the validity of  the Fr\"{o}hlich  Hamiltonian.) We finally remark that the condition for the occurrence of both of these characteristic motions within the Fr\"{o}hlich regime turns out to be $ \eta_{\rm max, d}<\eta_{\rm c, d}$. The latter are the critical couplings of  Fr\"{o}hlich regime given in the appendix \ref{App:DD} and correspond to the vertical lines shown in the Fig. \ref{fig:fig2}.

 \subsubsection{Average Energy}
We now turn to the average kinetic energy $E(t)$ of the impurity, corresponding to the observed coordinate.  This can be computed from the variance of the corresponding momentum operator reading
 \begin{align}
E_{x, \rm d}(t)=\frac{\langle p^{2}_{x, \rm d}(t)\rangle}{2 m_{\rm I}}
 \end{align}
The generalized expression for the variance of the momentum is given in the appendix \ref{App:CC}. Using the dimension-dependent asymptotic expressions for $G_{1}(t)$ and $G_{2}(t)$ we obtain
 \begin{align}
E_{x,\rm d}(t)=\frac{\langle p^{2}_{x, \rm d}(0)\rangle}{2 m_{\rm I}\pc{\alpha_{\rm d}}^2 }+\frac{\hbar}{2m_{\rm I}\pc{\alpha_{\rm d}}^2  }\!\!\int_{0}^{t}\!\!du\!\!\int_{0}^{t}\!\!dv~\nu^{xx}_{\rm d}(u-v).
 \end{align}
For any arbitrary temperature, it is difficult to obtain from here an analytic expression. Here we are mainly interested in the ultracold regime. This means that all the bath modes are now in a collective vacuum state. Therefore, in the zero-temperature limit, the above expression further reduces to
   \begin{align}
&E_{x,\rm d}^{\rm LT} (t)=\frac{\langle p^{2}(0)\rangle}{2 m_{\rm I}\pc{\alpha_{\rm d}}^2  }+\frac{\hbar \pc{\Lambda_{\rm d}} ^{d+1} \pc{ \tau_{\rm d}}^{d}}{\pc{\alpha_{\rm d}}^2 d (d+1)}+\nonumber\\
&\frac{\hbar \pc{\Lambda_{\rm d}}^{d+1}\pc{\tau_{\rm d}}^{d}}{\pc{\alpha_{\rm d}}^2 d (d+1)}
\left(\, _1F_2\left(\frac{d}{2}+\frac{1}{2};\frac{1}{2},\frac{d}{2}+\frac{3}{2};-\frac{1}{4} \pc{\Lambda_{\rm d}} ^2 t^2\right)\right).
 \end{align}
Here, the first term represents the initial mean energy of the impurity determined by its initial momentum variance. Additionally, there is a rescaling of the mass of the impurity due to the interaction with the bath. The additional mass term depends on the dimensionality of the bath through $ \Lambda_{\rm d}$ and $\tau_{\rm d}$. The second term is the steady state mean energy of the impurity which is determined by the impurity-bath coupling and density of the bath, again through the same parameters. The last term of the expression contains information about the energy variation in time.  We plot the energy function in Fig.  \ref{fig:fig3}~a-c) for different dimensions. In all dimensions, the energy oscillates in time.  This clearly shows the energy exchange between the system and the bath.  Where the energy increases, it is due to an energy absorbed from the bath.   The back flow of energy is a manifestation of memory effects in the QBM \cite{PhysRevA.93.012118}. Moreover, deep inside the weak coupling regime, the bath perturbs the system more strongly as the coupling strength increases. In any dimension, this results in the higher initial increasing peak for a larger coupling constant. The overall profiles of all the energy functions tend to approach their asymptotic steady state values.

A similar analysis can be performed for the high temperature case, as was done in the previous section for the MSD. We treat the problem classically, meaning that the the symmetrised noise correlation function (cf.  Eq.~ \eqref{symtricBSFCorrelation} and  Appendix \ref{App:CC} ) would act as the classical analogue in the present quantum formulation \cite{RevModPhys.82.1155}.  As for the dynamical part of the energy expression, the classical regime is further obtained by requiring the conditions $t\to \infty, \\ \hbar\to 0$ and $k_{\rm B}T\gg\hbar\omega$. In these limits,  the energy  $E_{x,\rm d}^{\rm cl, ss}$ becomes
 \begin{align}
E_{x,\rm d}^{\rm cl, ss}= 2k_{\rm B}T \pr{ \frac{ \beta_{\rm d}\eta^2}{\pc{\beta_{\rm d}}^{2}d^{-2}\eta^4+2\beta_{\rm d}\eta^2+d^{2}}}.
 \end{align}
The above expression is once again maximised at $ \eta_{\rm max, d}$ giving an upper bound to the kinetic energy of the impurity reading
 \begin{align}
E_{x,\rm d, max}^{\rm cl, ss}= k_{\rm B}T /2.
 \end{align}
Remarkably, this is the familiar equipartition theorem that holds in any dimension. It follows from these results that $ \eta_{\rm max, d}$ is the value of the system-bath coupling at which the impurity reaches thermal equilibrium with the Bogoliubov bath.

 \subsection{Trapped Case}
\begin{figure}[!t]
 \includegraphics[width=0.9\columnwidth]{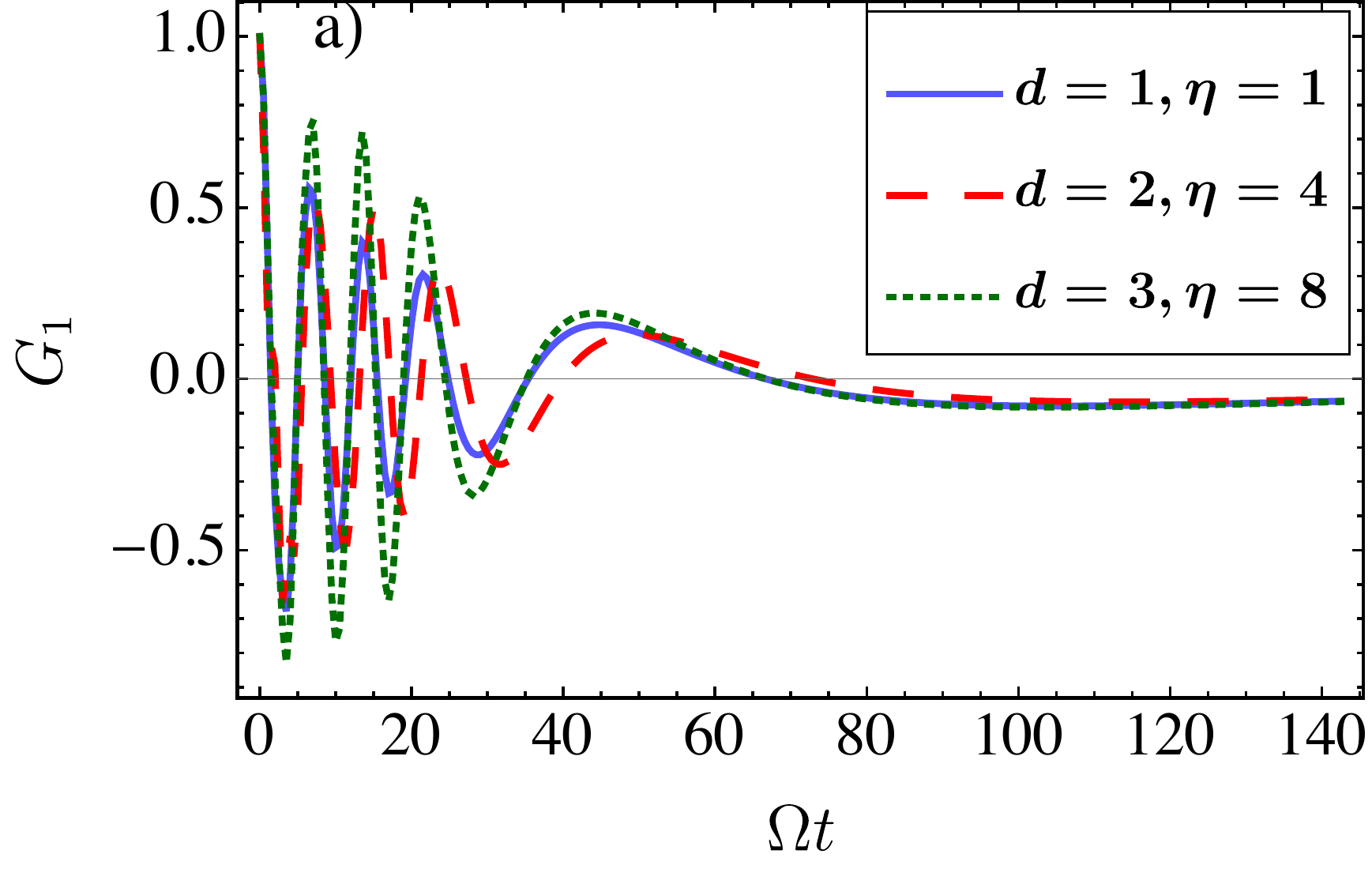}\\
 \includegraphics[width=0.9\columnwidth]{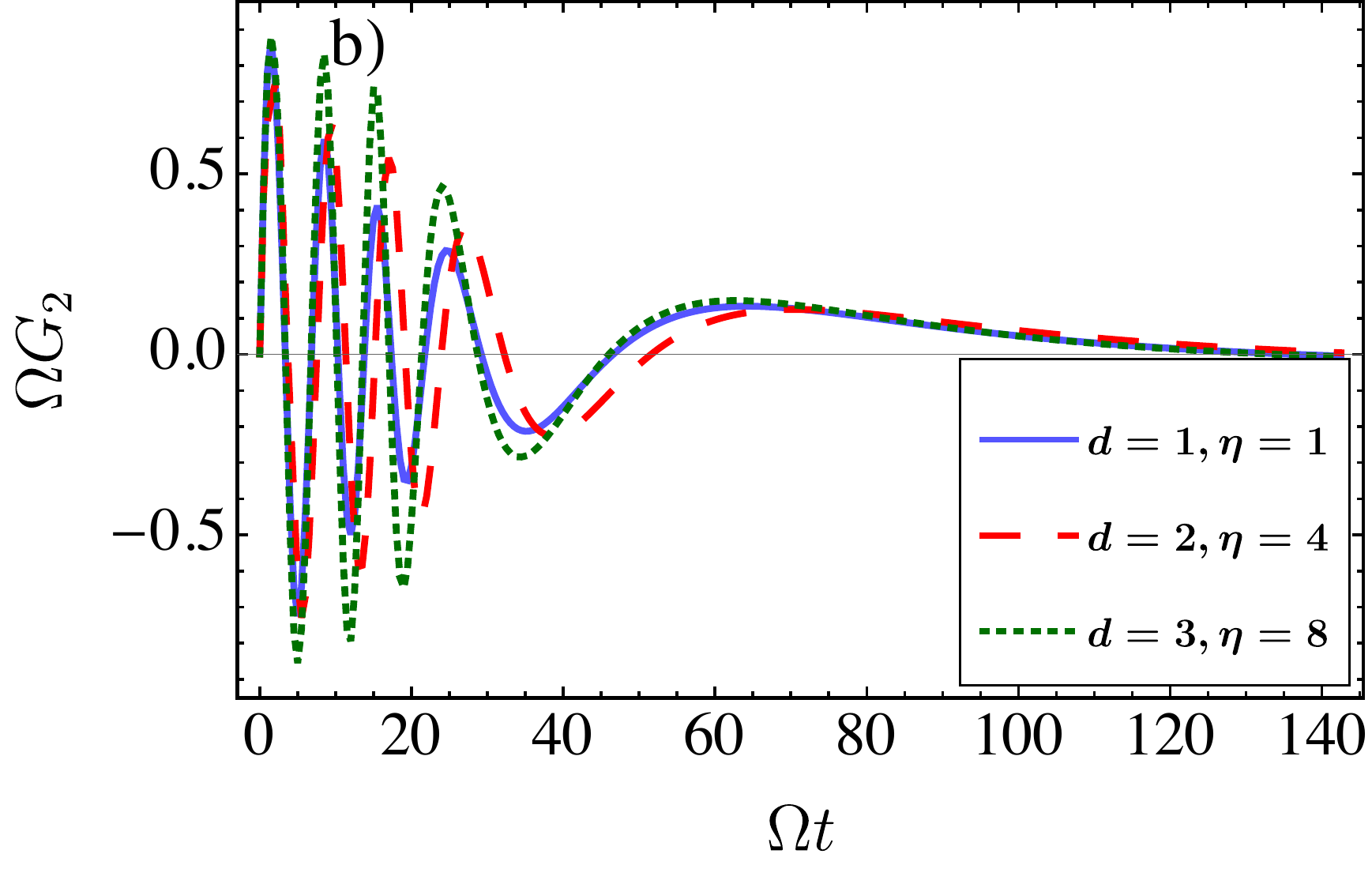}
\caption{ (a-b) Dynamics of $G_{1}(t)$ and $G_{2}(t)$ in the trapped case for 1$d$, 2$d$ and 3$d$ with the corresponding dimensions and couplings shown in the legend. In all the cases we have set $\Omega=4 \pi \times500~{\rm Hz}$. Rest of the parameters are same as in Fig.~\ref{fig:fig1}.}
 \label{fig:fig4}
\end{figure}

  In recent years, there has been an increased interest on trapped impurities within cold atomic media. For instance, the bound states of the trapped impurities provide a platform to test the existence of synthetic vacuum of the hosting medium by witnessing the induced Lamb-shifts \cite{Rentrop2016}.  Additionally, trapped impurities in BECs can serve as highly controlled phononic q-bits \cite{2013Scelle, PhysRevA.95.053618}. Hence, theoretical  study of a trapped impurity as an open quantum system, as in this work, can be valuable in all of the aforementioned cases. In the trapped case, we confine the impurity into a harmonic trap with frequency $\Omega$. We compute the functions $G_{1}(t)$ and $G_{2}(t)$ by employing the Zakian method. Their time dynamics is shown in Fig.~\ref{fig:fig4} (a-b). In all dimensions, both of these functions oscillate out of phase by $ \pi/2$. This reflects the fact that position and momentum are the two quadratures of the impurity motion. Note that the information of the initial position and the momentum variances is carried by the functions  $G_{1}(t)$ and $G_{2}(t)$ respectively (see  appendix \ref{App:CC}).  The decay of these functions provides insights into the system dynamics. First  of all,  such decay shows that the impurity dynamics is stable. In general, the stability analysis of the dynamics can be performed more rigorously, e.g. through the Routh–Hurwitz stability criterion \cite{Morris1962}. However, given the absence of the analytical form of $G_{1}(t)$ and $G_{2}(t)$ (or their Laplace transforms), we rely on a numerical evaluation of their profiles.
In fact, both of these functions approach to zero as $t\rightarrow\infty$. In effect, the system dynamics becomes independent of its initial conditions and its behaviour is completely determined by the bath state. It can be seen that in the long time limit, each one of them collapses to a single  curve for all the cases displayed in Fig. 4.   On the contrary at initial short times, their amplitudes and phases are mismatched for different initial conditions. The  differences coming from different coupling parameters and dimensions also vanish in the long time limit. This leads to the insight that the steady state regime is completely determined by the bath state and the system tends to equilibrate with the local state of the bath. 

We now turn to the study of the steady state dynamics of the impurity. Since the input Bogoliubov bath modes are in a Gaussian thermal state, and the linear dynamical map \eqref{eq:EOMPositionImpurityXdirection} preserves Gaussianity, the time evolution of the covariance matrix fully characterises the impurity dynamics.  Here we are interested in the position and momentum variances. Their expressions for the case of generalised Langevin equations are given in appendix \ref{App:CC}. In  particular, we are interested in the position variance of the impurity in the steady state regime as a function of the dimension and other parameters such as temperature of the bath. This is because in larger dimensions, the impurity will have more  degrees of freedom (dof) and one wants to see how the energy is distributed among them in the steady state. This is particularly interesting in the finite temperate case, where the system is more prone to achieve a thermal equilibrium with the bath.

 \begin{figure*}[]
 \includegraphics[width=0.68\columnwidth]{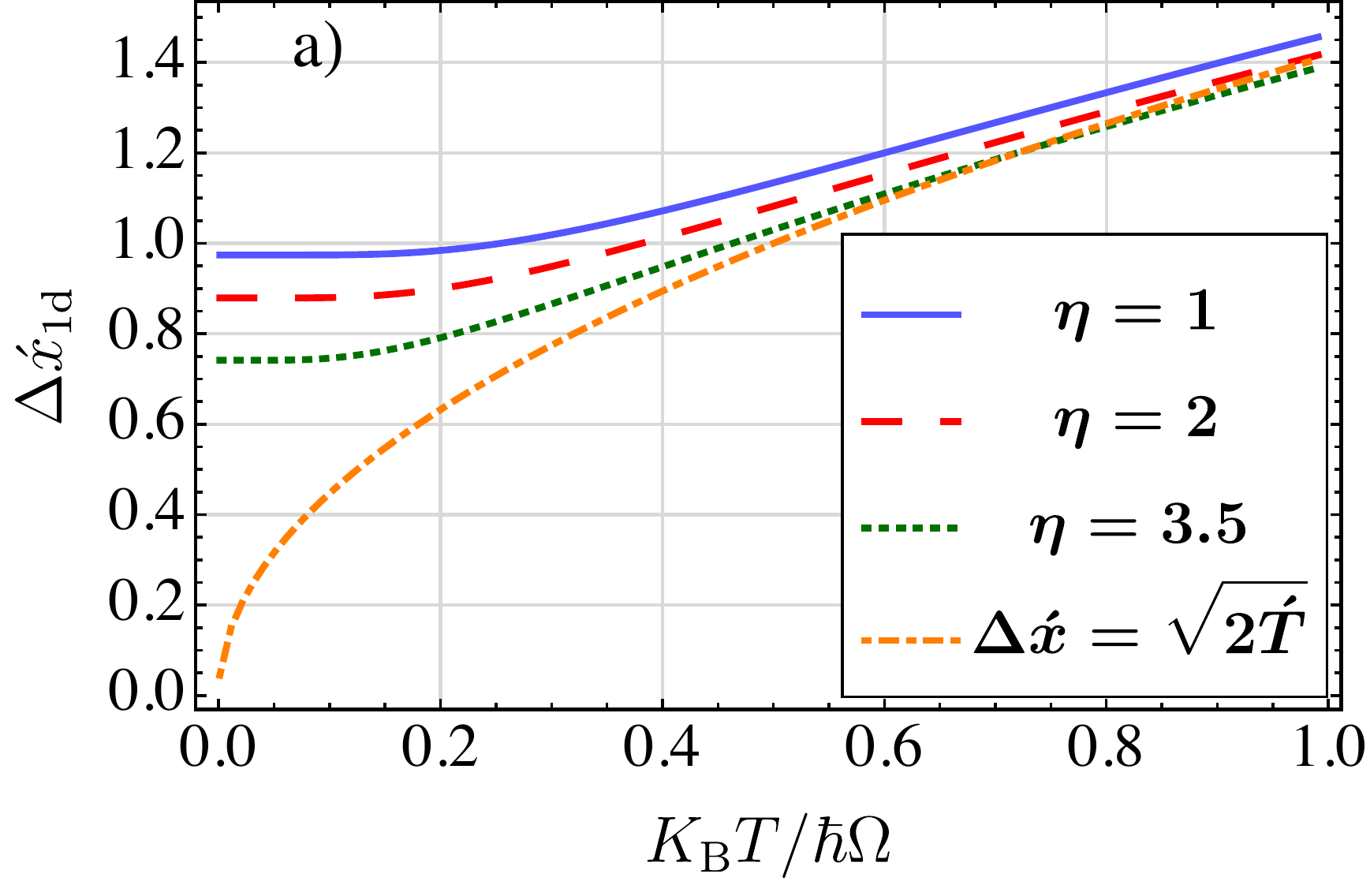}
 \includegraphics[width=0.68\columnwidth]{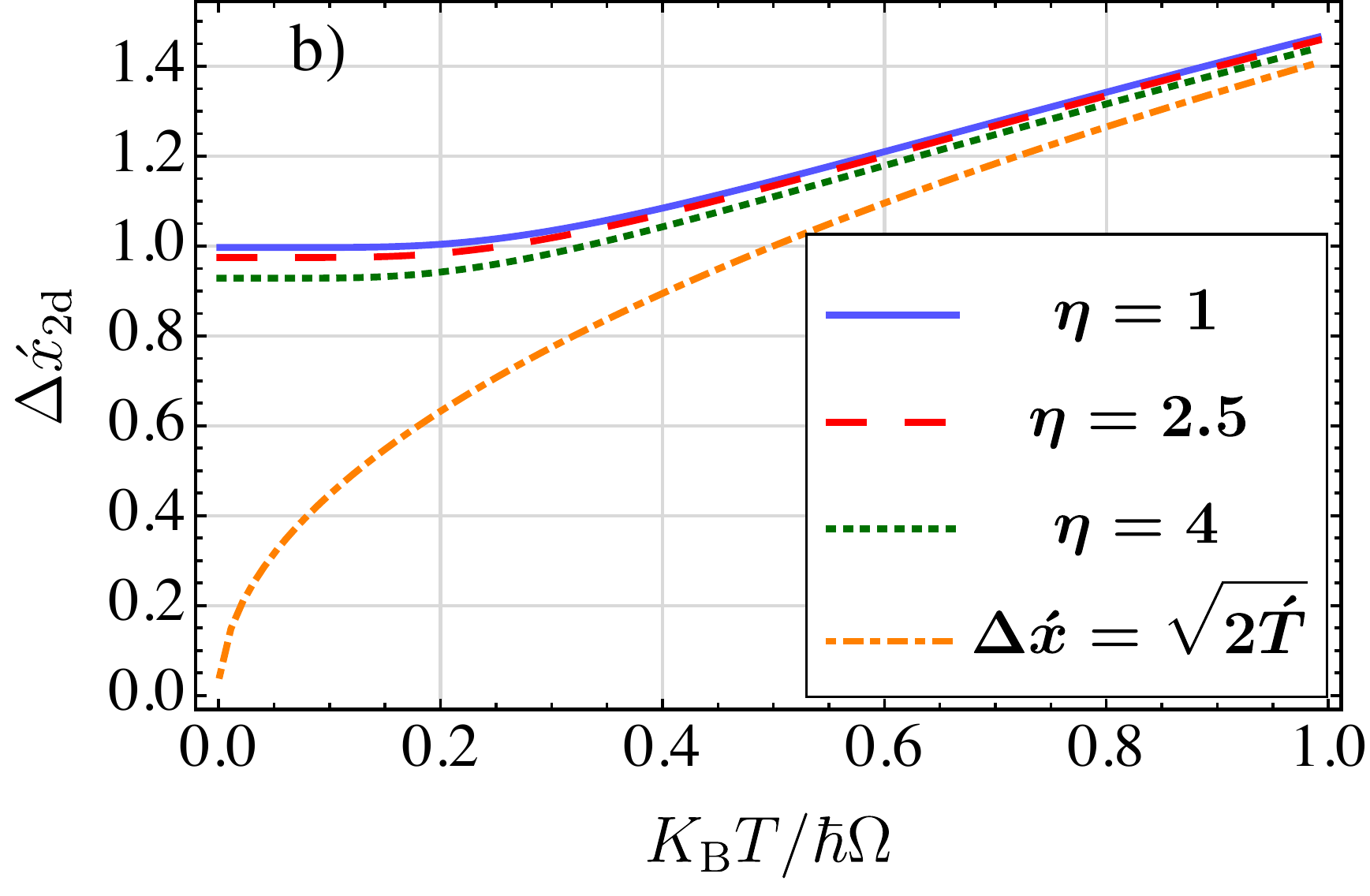}
\includegraphics[width=0.68\columnwidth]{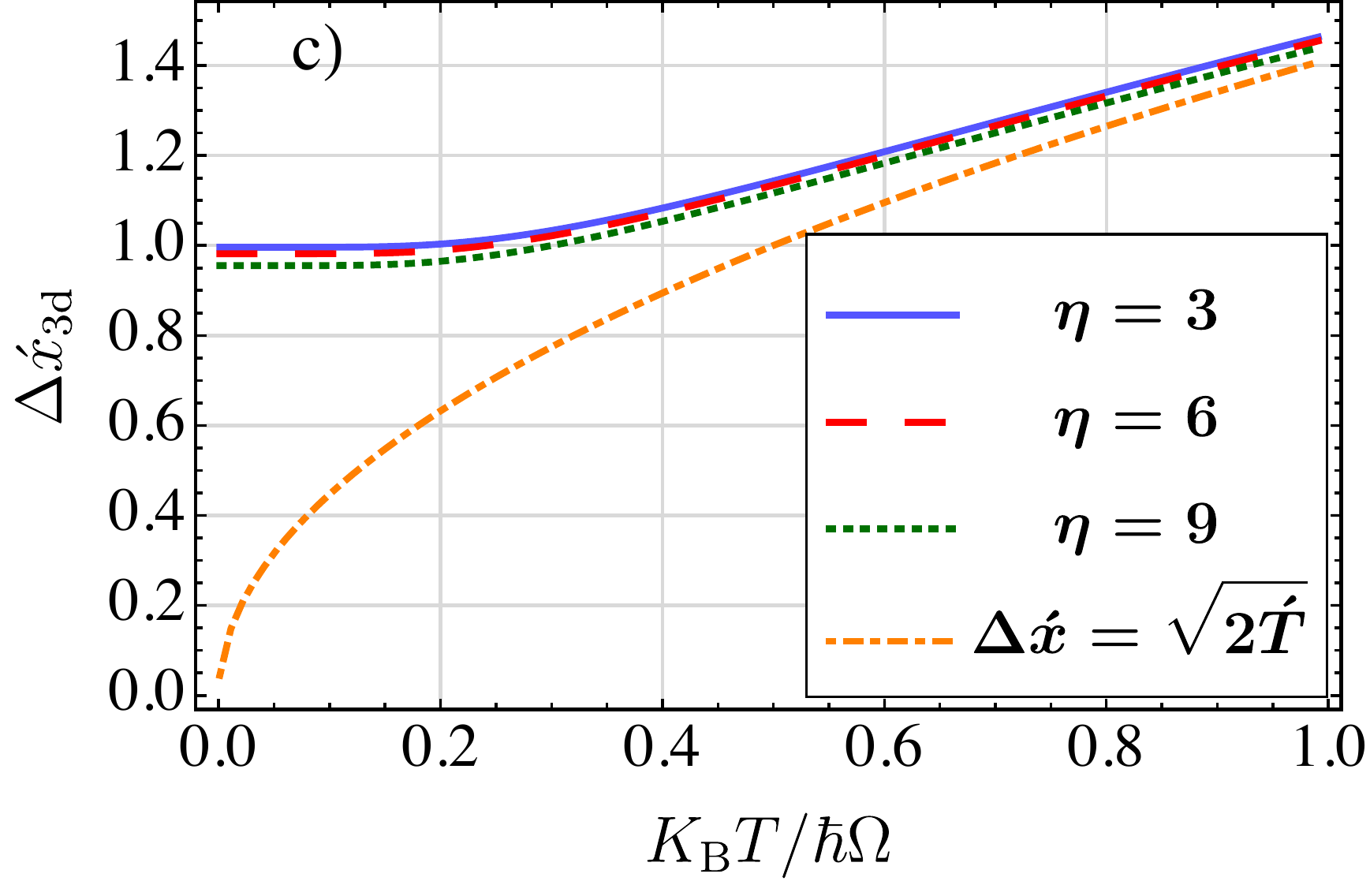}
  \caption{ (a-c) Steady state position squeezing of the impurity as a function of temperature for the case of 1$d$, 2$d$, and 3$d$, respectively.}
   \label{cfig:fig5}
\end{figure*}
We first calculate the position variances for the trapped case in the steady state regime.   We start from its expression given in the appendix \ref{App:CC}. By invoking the component of noise tensor which is responsible for the $x$-directed motion and in the long time limit, for the $d$-dimensional system we have
\begin{align}
\av{x^2}_{\rm d}=\lim_{t\rightarrow\infty}\frac{\hbar}{m_{\rm I}^2}\int_{0}^{t}du\int_{0}^{t}dv& G_{\rm 2,d}(t-u) G_{\rm 2,d}(t-v)\nonumber\\
&\times \nu_{\rm d}^{xx}(u-v).
\label{eq:positionvarienSteadystate}
 \end{align}
This can be further written as
\begin{align}
&\av{x^2}_{\rm d}=\lim_{t\rightarrow\infty}\frac{\hbar}{m_{\rm I}^2}\int_{0}^{\infty}d\omega J_{\rm d}^{xx}\pc{\omega}\coth\pc{\hbar\omega/2k_{\rm B}T}\nonumber\\
&\int_{0}^{t}du\int_{0}^{t}dv G_{\rm 2,d}(t-u) G_{\rm 2,d}(t-v)\cos\pr{\omega\pc{u-v}}.
\label{eq:positionvarienSteadystate2}
 \end{align}
We now define the collective function  $\mathcal{Q}_{\rm d}(\omega)$ 
\begin{align}
\mathcal{Q_{\rm d}}(\omega)&\equiv\lim_{t\rightarrow\infty}\pc{1/2}\pc{\frac{1}{m_{\rm I}^2}}\int_{0}^{t}du\int_{0}^{t}dv\nonumber\\
&\times G_{\rm 2,d}(t-u) G_{\rm 2,d}(t-v)\pr{ e^{i\omega u}e^{-i\omega v}+c.c.}\nonumber\\
&=\pc{\frac{1}{m_{\rm I}^2}}\pc{1/2}\lim_{t\rightarrow\infty}\int_{0}^{t}d\tilde{u}e^{-i\omega \tilde{u}}G_{\rm 2,d}(\tilde{u})\int_{0}^{t}d\tilde{v}e^{i\omega \tilde{v}}\nonumber\\
&\quad \quad G_{\rm 2,d}(\tilde{v})+c.c.
\label{eq:Qfunction1}
 \end{align}
 so that
\begin{align}
\av{x^2}_{\rm d}=\hbar\int_{0}^{\infty}d\omega J_{\rm d}^{xx}\pc{\omega}\coth\pc{\hbar\omega/2k_{\rm B}T}
\mathcal{Q_{\rm d}}(\omega).
\label{eq:positionvarienSteadystate3}
 \end{align}
In the second equality of Eq. \eqref{eq:Qfunction1}, we have defined new variables $\tilde{u}=t-u $ and $\tilde{v}=t-v$. It turns out that each term is the product of two copies of the Laplace transform of $G_{\rm 2,d}(\tilde{u})$, evaluated at $\mathcal{S}=-i\omega$ and $\mathcal{S}=i\omega$. We therefore obtain
\begin{align}
\mathcal{Q_{\rm d}}(\omega)&=\pc{\frac{1}{m_{\rm I}^2}}\mathcal{L}_{\mathcal{S}=-i\omega, \rm d}\pr{G_{\rm 2,d}(t)}\mathcal{L}_{\mathcal{S}=i\omega,\rm d}\pr{G_{\rm 2,d}(t)}.
\label{eq:Qfunction2}
\end{align}
We now introduce the frequency dependent functions $ \xi^{xx}_{\rm d}(\omega)\equiv\Re\px{ \mathcal{L}_{\bar{\mathcal{S}},{\rm d}}\pr{ \Gamma_{\rm d}^{xx}\pc{t}}}$ and $ \theta^{xx}_{\rm d}(\omega)\equiv\Im\px{ \mathcal{L}_{\bar{\mathcal{S}},{\rm d}}\pr{ \Gamma\pc{t}}}$ which are respectively  the real and imaginary parts of the Fourier domain tensor component of the damping kernel. They are obtained by the analytic continuation of Eq. (\ref{eq:LaplaceDamping}) with $\bar{\mathcal{S}}=-i \omega+0^{+} $. This allows us to write the function in  Eq. \eqref{eq:Qfunction2} as 
\begin{align}
\mathcal{Q_{\rm d}}(\omega)&=\pc{\frac{1}{m_{\rm I}^2}}\pc{\frac{1}{\pr{\Omega^{2}-\omega^{2}-\omega\theta^{xx}_{\rm d}(\omega)}^{2}+\pr{\omega\xi^{xx}_{\rm d}(\omega)}^{2}}}.
\label{eq:Qfunction3}
\end{align}
Additionally, from the relation between the damping kernel and the spectral density tensor components given in the Eq. \eqref{eq:DampingKernal}, one can derive the equation\cite{Weiss2008}
\begin{align}
J_{\rm d}^{xx}(\omega)=\frac{m_{\rm I}\omega\xi^{xx}_{\rm d}(\omega)}{\pi}.
\label{eq:spectralfunctionOFxi}
\end{align}
By inserting Eq. \eqref{eq:Qfunction3} and Eq. \eqref{eq:spectralfunctionOFxi} back into the Eq. \eqref{eq:positionvarienSteadystate3}, we finally obtain

\begin{align}
\av{x^2}_{\rm d}=\frac{\hbar}{\pi}\int_{0}^{\infty}d\omega\coth\pc{\frac{\hbar\omega}{2k_{\rm B}T}}\tilde{\chi}_{\rm d}^{''}(\omega),
\label{eq:SteadyStatePositionVariene}
\end{align}
where  the function  $\tilde{\chi}_{\rm d}^{''}(\omega)$ equals
\begin{align}
\tilde{\chi}_{\rm d}^{''}(\omega)=\frac{1}{m_{\rm I}}\frac{\xi_{\rm d}^{xx}(\omega) \omega}{\pr{\omega\xi_{\rm d}^{xx}(\omega)}^{2}+\pr{\Omega^{2}-\omega^{2}+\omega \theta_{\rm d}^{xx}\pc{\omega}}^{2}}.
 \end{align}
From the direct inspection of Eq. \eqref{eq:SolutionEOM}, it turns out that this function is the imaginary part of the susceptibility  $\tilde{\chi}_{\rm d}=\tilde{\chi}_{\rm d}^{'}+ i \tilde{\chi}_{\rm d}^{''}$. The susceptibility function can be obtained by extracting the linear response function from Eq. \eqref{eq:SolutionEOM} in the Laplace domain.  We can continue analytically to pass to the frequency representation (i.e. Fourier domain).   Note  that the function  defined above is nothing but the absolute square of the susceptibility i.e. $|\tilde{\chi}_{\rm d}(\omega)|^2 =\mathcal{Q_{\rm d}}(\omega)$. 

It follows from these results that the steady state position variance is fully determined by the  functions $ \xi^{xx}_{\rm d}(\omega)$ and  $ \theta^{xx}_{\rm d}(\omega)$.  In addition,  note that the upper limit of the Eq. \eqref{eq:SteadyStatePositionVariene} is reduced to the cut-off frequency in every dimension due to the unit step function involved in spectral function. We perform an analytic continuation of  Eq. (\ref{eq:LaplaceDamping}) to get these functions in the low frequency limit $\omega\ll\Lambda_{\rm d}$  (up to first few orders of $\omega$) for each of the dimensions.  Such a process is straightforward for $d=1$ and $d=3$, since both can be expanded as polynomials of $\omega$. This is not the case for $d=2$, since in this case the expression  contains a logarithmic transcendental function of the frequency. In order to study this function for low frequencies, we evaluate it numerically below the cut-off, and use this for a numerical evaluation of the position variance. We pass to the dimensionless (scaled) position quadrature by introducing $\acute{x}\equiv x/x_{zpf}$, where $x_{zpf}\equiv\sqrt{\hbar/ 2m_{\rm I}\Omega}$ is the zero-point fluctuation (ground state width) of the harmonically bound impurity. The Heisenberg uncertainty relation, for the standard deviations of the conjugate position and momentum operators,  in the scaled variables becomes
\begin{align}
\Delta \acute{x}_{\rm d}\Delta \acute{p}_{\acute{x}, \rm d}\geqslant 1.
 \end{align}
On the other hand, if one of these standard deviations  falls below unity, it is said to have achieved a squeezed state. The squeezing is a pure quantum effect where quantum noise is driven below its ground state uncertainty for one of the conjugate observables. The mechanical squeezed states are of great significance in high precision displacement sensing \cite{Wollman952}. 
One may also express such standard deviations in the scaled coordinates. For the position and momentum, we get
\begin{align}
\Delta \acute{x}_{\rm d}=\sqrt{\frac{2 m_{\rm I}\Omega \av{x^2}_{\rm d}}{\hbar}}, \quad \Delta \acute{p}_{\acute{x}, \rm d}=\sqrt{\frac{2  \av{p_{x}^2}_{\rm d}}{\hbar m_{\rm I}\Omega}}.
 \end{align}
From here on, we focus on the position quadrature. Before proceeding to study position squeezing, we comment on the equipartition theorem in our system. In the present case, we have a thermostat with the large number of modes of the Bogoliubov bosonic bath. Additionally, one can tune the temperature of the bath sufficiently high for all the modes to be thermally populated. In the long time limit, the bath achieves a thermal equilibrium steady state. An immersed impurity would therefore tend to equilibrate too, once the temperature of the bath is kept above (or close to) its trap frequency. In thermal equilibrium, equipartition theorem states that the amount of energy $(1/2)k_{\rm B}T$ is distributed per degree of freedom. We thus have
\begin{align}
(1/2)k_{\rm B}T=(1/2)m_{\rm I}\Omega^{2}\av{x^2}_{\rm d}.
 \end{align}
and thus the standard deviation in the dimensionless coordinate reads
\begin{align}
\Delta \acute{x}_{\rm d}=\sqrt{2 \acute{T}}, \quad \text{where} \quad \acute{T}\equiv(k_{\rm B}T)/(\hbar\Omega).
\end{align}
By checking that the position squeezing parameter asymptotically approaches the equipartition profile as stated in the last equation, one can verify that impurity motion follows the equipartition theorem. This is indeed the case, as shown in  Fig. \ref{cfig:fig5} (a-c). As the thermal energy $k_{\rm B}T$ becomes equal to the quantum energy $\hbar\Omega$, all the cases approach the equipartition profiles. Note that this holds in all dimensions $d = 1, 2, 3$, despite the fact that the tensor component of the bath spectral density $J^{xx}$, which is responsible for the $x$-directed motion,  scales as $d^{-1}$ [cf.  Eq. \eqref{eq:JTensorMainText}].

On the other hand, the differences between the profiles corresponding to different dimensions are apparent when one examines the magnitude of the position squeezing achieved. Although a direct comparison is not possible due to different ranges of coupling compatible with the Fr\"{o}hlich regime, it is apparent that squeezing is more pronounced in lower dimensions.  The amount of squeezing is proportional to the coupling strength and it is achieved at quite low temperature. The squeezing effect is purely due the interaction with the bath and it occurs without an external control of the impunity motion. It is referred to as genuine position squeezing \cite{2017Lampo}.

\section{Memory Effects}\label{sec:memory}
\begin{figure}[!t]
 \includegraphics[width=0.8\columnwidth]{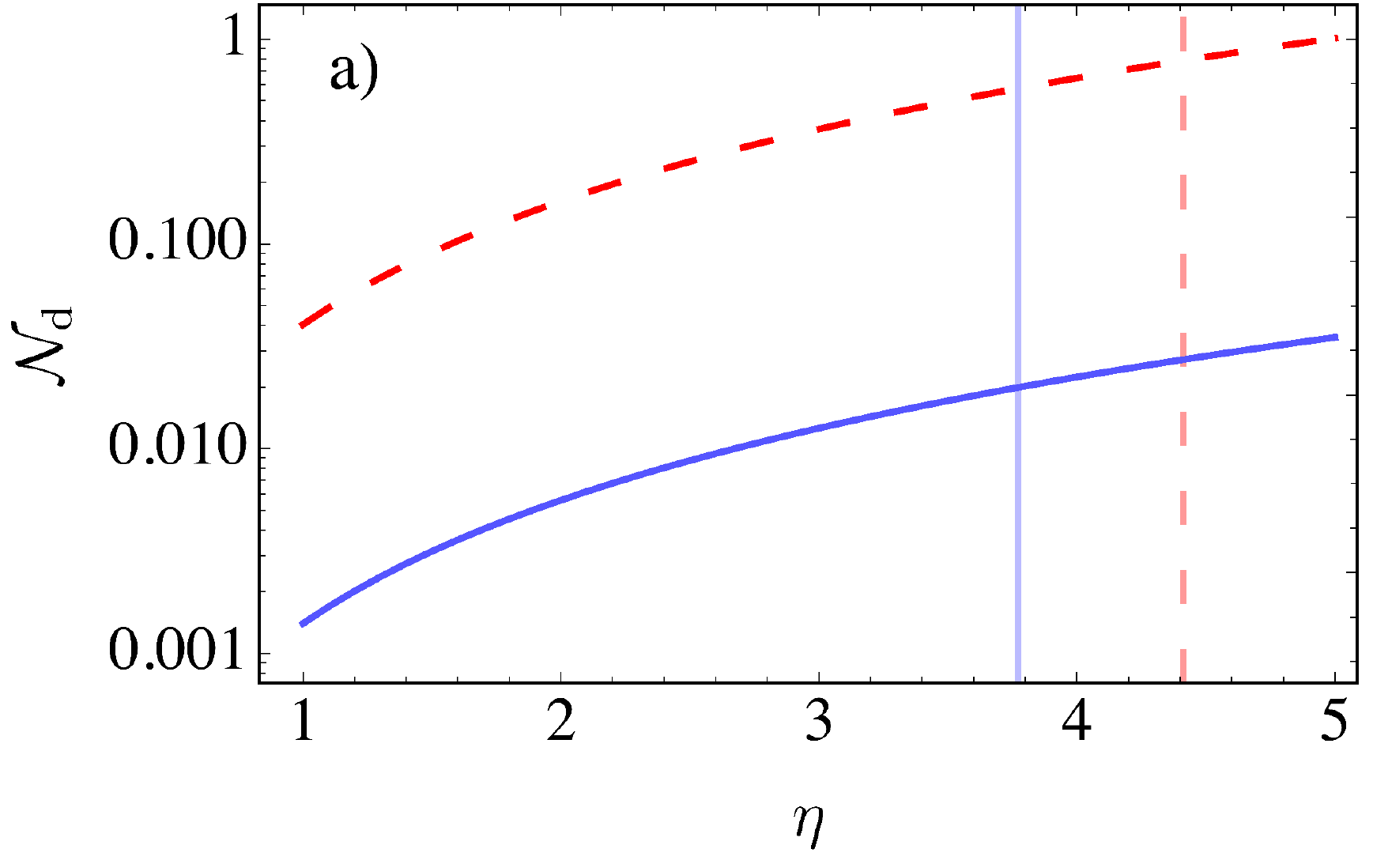}\\
 \includegraphics[width=0.8\columnwidth]{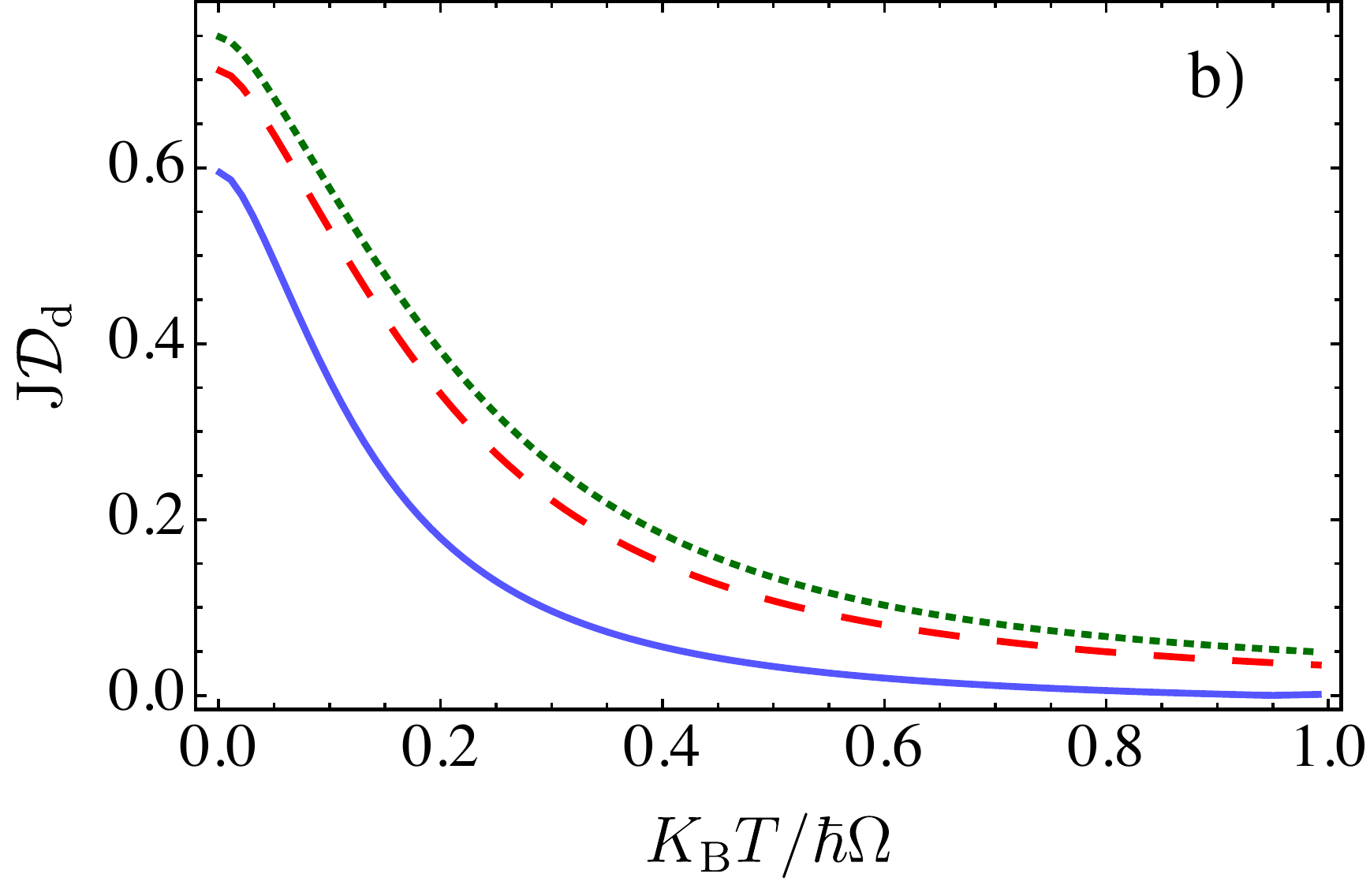}
\caption{a) Non-Markovianity in units of $\pc{g_{\rm B, 1}~ n_{0 \rm, 1}~m_{\rm I}}/\hbar$ as a function of the system-bath coupling strength. The blue-Solid and red-dashed profiles represent  $1d$ and $2d$, respectively. The vertical lines correspond to the Fr\"{o}hlich bound.  b)  The $J$-distance as a measure of the relative position variance. We have set the coupling parameter $\eta=3.5$ and  $\gamma=10~\Omega$ for every dimension. Here, the solid, dashed and dotted profiles correspond to $1d$, $2d$ and $3d$, respectively.}
 \label{fig:fig6-7}
\end{figure}
In this section, we briefly analyse  the memory effects. In general, one expects to see Makovian dynamics when the spectral density of the bath is Ohmic, and for sufficiently high temperature and weak coupling, such as that obtained with a Dirac delta like damping kernel in Eq. \eqref{eq:EOMPositionImpurityXdirection}. On the contrary, the super-Ohmic nature of the bath is generally assumed to lead to non-Markovian dynamics of the system, i.e. to  memory effects.  We will use the super-Ohmic spectral densities of the bath we obtained for different dimensions to study the relation between the expected non-Markovian effects and dimensionality. In order to quantify the amount of non-Markovianity,  a number of measures have been proposed so far \cite{RevModPhys.88.021002}.  Here, we apply the criteria related to the back flow of information. It has been shown that such back flow of information can be expressed in terms of the trace distance \cite{PhysRevLett.103.210401} and the fidelity \cite{PhysRevA.84.052118} of two states. For Gaussian states, the latter has an analytical form. More specifically, the non-Markovianity  $ \mathcal{N}_{\rm d}$ based on the back flow of information through the fidelity criterion is explicitly related to the noise kernel and is given by
\begin{align}
\mathcal{N}_{\rm d}=\int_{\Delta_{\rm d}<0}\Delta_{\rm d}(t)dt \quad \text{with}
 \end{align}
\begin{align}
\Delta_{\rm d}(t)=\int_{0}^{t}\nu^{xx}_{\rm d}(s) \cos(\Omega s).
 \end{align}
In our case, the noise kernel is given by Eq. (\ref{eq:NoiseKernal}) which depends on the dimension of the system. The above measure can be obtained by first calculating the definite integrals over the variables $\omega$   and $s$, and then integrating the resulting function $\Delta_{\rm d}(t)$ over the time region where it is negative. In the study of non-Markovianity through this measure, we focus on the comparison between the cases of $d=1$ and $d=2$ dimensions. In Fig. \ref{fig:fig6-7} a), we plot the measure $\mathcal{N}_{\rm d}$ for the zero temperature case (i.e. when the cotangent in Eq.~ (\ref{eq:NoiseKernal}) equals $1$) for a range of values of the system-bath coupling.  Clearly the non-Markovianity measure is showing a monotonically increasing behaviour on the logarithmic scale, as we enhance the system-bath coupling. In addition, as we move from low to higher dimension with the corresponding higher level of super-Ohmicity, larger  non-Markovian effects are witnessed:  the non-Markovianity is at least an order of magnitude larger in $d=2$ than in  $d=1$. 

Another measure of memory effect is the $J$-distance $\rm J \mathcal{D}$, which has been introduced in Ref.  \cite{2018Lampo}. This is a quantitative measure of  the relative position variance between the case where the spectral density is super-Ohmic and the case where it is Ohmic. The Ohmic case corresponds to the time-local form of the damping kernel and has spectral density $J_{\rm Ohm}=m_{\rm I}\gamma \omega$, where we have introduced the phenomenological damping constant $\gamma$.  We may explicitly write the $J$-distance as
\begin{align}
\rm J \mathcal{D}_{\rm d}=\left|\frac{\langle x^{2}\rangle_{\rm d}-\langle x^{2}\rangle_{\rm Ohm, d}}{\langle x^{2}\rangle_{\rm d}+\langle x^{2}\rangle_{\rm Ohm, d}}\right|.
 \label{Jdistmeasure}
 \end{align}

Here  $\langle x^{2}\rangle_{\rm Ohm, d}$ is the position variance that would be obtained, had we assumed an Ohmic Spectral density with the same cut-off function. In Fig. \ref{fig:fig6-7} (b), we plot this  quantity for different dimensions as a function of temperature.  It turns out that for a similar system-bath coupling strength $\rm J \mathcal{D}_{\rm d}$ assumes largest values for $d=3$. This is supported by the argument that in a higher dimension super-ohmicity affects the dynamics stronger, leading to a larger deviation from the ohmic case.   In addition, the non-Markovian effects are more pronounced near zero temperature  i.e.  a vacuum bath. As the temperature grows, dynamics tends to achieve the thermal steady state both for  Markovian and non-Markovian scenarios. The relative difference in Eqn.~\eqref{Jdistmeasure} then approaches zero.
 
\section{Conclusions}\label{sec:conclusions}

In this work, we studied from the quantum open systems perspective the dynamics of an impurity immersed in a $d$-dimensional BEC. In particular, we offered a detailed derivation of the Langevin equations and the associated generalized $d$-dimensional spectral density. We derived an expression for the tensor that describes this spectral density in full generality. Particular attention was given to the case of a spherically symmetric bath, which implied a diagonal form for this tensor. In addition, the tensors for the noise and damping kernels were calculated; these tensors enter  the vectorial Langevin-like  equations of motion. All these technical aspects, allowed to study in detail the dependence of the dynamics of the impurity on the dimensionality.  

We considered both untrapped and trapped scenarios for the impurity.  In the untrapped case, we  performed in all dimensions the calculation of the mean square displacement, showing that the motion is superdiffusive.  We derived explicit expressions for the superdiffusivity coefficient in the low and high temperature limits. In the latter limit we found that this coefficient has a maximum as a function of the impurity-boson coupling. The maximal value of the coefficient is equal in all dimensions, but the value of the coupling at which it occurs is dimension-dependent.  It lies within the limits of validity of the Fr\"ohlich model in one dimension only.  We calculated also the average energy, for which we obtained the generalized expression for the variance of the momentum;  the expression of average energy in the ultracold regime was calculated explicitly. These results confirm the expected rescaling of the mass of the impurity as a consequence of its interaction with the bath, whose specific value depends on the dimensionality. The behaviour of the energy in all dimensions is oscillatory, with a backflow of the energy between the bath and the impurity---a memory effect. We performed a similar analysis for the large temperature limit, finding the expressions for very large times and in the classical limit. These expressions exhibit  maximum at certain value of the impurity-boson coupling and, amazingly, at this value of the coupling, the equipartition theorem is fulfilled in all dimensions, so the impurity is in a thermal equilibrium with the Bogoliubov bath.

In the trapped case, we obtained the steady state and characterized it with its covariance matrix. To this end, we obtained  the expressions of the position and momentum variances. We identified the position variance is connected to the the imaginary part of the susceptibility. To calculate the position variance explicitly is possible in one and three dimensions, but in two dimensions it involves a logarithmic transcendental function of the frequency, so we proceeded numerically. We saw that one can find squeezing in all dimensions, which is an important result as it can be used for applications in  quantum technologies, such as quantum sensing and metrology. Here we evaluated it in any dimensions, which is relevant for many experimental set-ups. In the high temperature limit we calculated the equipartition profile and showed that in all dimensions the variances tend to this limit as temperature is increased. We found that, although a direct comparison among dimensions is not possible within our framework, due to different ranges of  admissible coupling strengths within the  Fr\"{o}hlich regime,  the magnitude of the position squeezing achieved is different, and at  low dimensions one can obtain stronger position squeezing. 

Finally, we also computed the amount of non-Markovianity in all dimensions via two quantifiers: the backflow of energy and the, so called, $J$-distance. Again, the direct comparison among dimensions is not possible, due to different parameter regimes. For the back flow of energy, we only perform the calculation in one and two dimensions, as for increased dimensions more and more cumbersome functions appear in the expressions, which complicate the analysis. Nevertheless this suffices to show that non-Markovianity grows with interactions and increases with dimensionality. The latter effect is also apparent in the calculation of the $J$-distance, which shows that the  non-Markovian effects are sizeable in higher dimensions. As an outlook, we foresee that this work will permit to continue the explorations of the Bose polaron problem in different quantum technologies, where the role of squeezed states and of non-Markovian effects may be important.  Finally, we also hope that in the future, this work will provide the starting point for investigations of more complex trapping settings, both for the impurity as well as for the bath itself.

\section*{Acknowledgments}
We thank A. Lampo for encouraging and motivating discussions. 
 M.K. and H.T. acknowledge support from Funda\c{c}\~ao para a Ci\^encia e Tecnologia (FCT-Portugal) through Grant No PD/BD/114345/2016 and through Contract No IF/00433/2015 respectively. H.T. and J.T.M. acknowledge the financial support from the Quantum Flagship Grant PhoQuS (Grant No. 820392) of the European Union. M.A.G.M. acknowledges funding from the Spanish Ministry of Education and VocationalTraining (MEFP) through the Beatriz Galindo program 2018 (BEAGAL18/00203).  We (M.L. group) acknowledge the Spanish Ministry MINECO (National Plan 15 Grant: FISICATEAMO No. FIS2016-79508-P, SEVERO OCHOA No. SEV-2015-0522, FPI), European Social Fund, Fundació Cellex, Fundació Mir-Puig, Generalitat de Catalunya (AGAUR Grant No. 2017 SGR 1341, CERCA program, 
 QuantumCAT$\_$U16-011424 , co-funded by ERDF Operational Program of Catalonia 2014-2020), MINECO-EU QUANTERA MAQS (funded by The State Research Agency (AEI) PCI2019-111828-2 / 10.13039/501100011033) , and the National Science Centre, Poland-Symfonia Grant No. 2016/20/W/ST4/00314.   J.W. was partially supported by the U.S. National Science Foundation grant DMS 1615045.
 
\appendix

\section{Spectral density tensor} 
\label{App:AA}
In this appendix we derive the spectral density tensor by employing the Bogoliubov dispersion relation.  We start from the form of the coupling  $\vect{g}_\mathbf{k}$ given by Eq. \eqref{eq:CouplingParameterSytemBath} and the definition of the spectral density tensor in Eq. \eqref{eq:Spectral DensityDiscrete1}. In general, the large number of the oscillators within the Bogolibouv bath allows us a continuous variable to label modes in  the $\mathbf {k}$-space.   In the  following, we use matrix representation of all the vectors (or, more generally, tensors). We perform the mathematical construction for each  dimension in the $\mathbf {k}$ space separately.

\textit{\textbf{3d Bath}}: 
A  three dimensional $k$-space representing the Bogolibouv bath is spanned by $k_{x}, k_{y}$ and $ k_{z}$. We introduce the polar angle $\phi $ (angle with $k_{z}$ direction ) and azimuthal angle $\theta$  (angle with $k_{x}$ direction) in the $k$-space. By employing  Eq.~\eqref{eq:CouplingParameterSytemBath}, the vectorial coupling is written in these angles as
 \begin{align}
&\underline{g_{k}}=(1/\hbar) \nonumber\\
&\pr{V_{k}k \cos (\theta ) \sin (\phi ),V_{ k}k \sin (\theta ) \sin (\phi ) ,V_{ k}k \cos (\phi )  }^{ T}.
 \end{align}
The coupling tensor is given by $ \underline{\underline{g_{k}}}\equiv \underline{g_{k}}~ \underline{g_{k}}^{T} $. Here $\pr{.}^{T}$ denotes the transpose conjugation. 
The spectral density tensor is 
\begin{align}
\underline{\underline{J}}(\omega)=\sum_{k}\hbar\underline{\underline{g_{k}}}\delta\pc{\omega-\omega_{k}}.
 \end{align}
In the continuum limit, we perform the transformation
 \begin{align}
\sum_{k}\rightarrow \int \frac{V}{(2\pi)^{3}} d^{3}k.
 \label{eq:}
 \end{align}
 Integrating over $dk\times kd\phi\times  k\sin{\phi}d\theta$ with appropriate limits, we obtain the explicit expression of the spectral density tensor $\underline{\underline{J_{3}}}(\omega)$ in the $3d$ case:
 \begin{align}
\underline{\underline{J_{3}}}(\omega)=&\frac{V}{(2\pi)^3}\int_{0}^\infty  dk \int_{0}^{\pi }~kd\phi \int_{0}^{2\pi }k\sin{\phi}d\theta\times\nonumber\\
&\pc{\hbar ~\underline{\underline{g_{k}}}~ \sum_{k_{\omega}}\frac{1}{\partial_{k}\omega_{k}|_{k=k_{\omega}}}\delta\pc{k-k_{\omega}}}.
\label{AA:SpctralDensityTensor}
 \end{align}
Here we have replaced the delta function with the $\omega$  argument by one with the $k$ argument, inverting the relation between these variables, as given by the dispersion relation Eq.~\eqref{eq:BogliubovSpec}. Moreover, the $k_{\omega}$ are the roots of the argument of the former delta function  i.e. of the equation $\omega-\omega_{k}=0$. A calculation shows that the contributing real root is  $k_{\omega}=\xi^{-1}(\sqrt{1+2 (\xi \omega/c)^{2} }-1)^{1/2}$.
 
 \textit{\textbf{2d Bath}}:
We parametrize the two dimensional $k$-space by the azimuthal angle $ \theta$ and the radial vector amplitude $k$. We  define in this case the two dimensional vector
 \begin{align}
\underline{g_{k}}=(1/\hbar) \pr{V_{k}k \cos (\theta ),~V_{k}k \sin (\theta ) }^{ T}.
 \end{align}
The calculation of the spectral density tensor is similar as in the previous case; the difference is that here we calcuate the integrals
 \begin{align}
\underline{\underline{J_{2}}}(\omega)=&\frac{V}{(2\pi)^2}\int_{0}^\infty  dk \int_{0}^{2\pi }kd\theta\times \nonumber\\
&\pc{\hbar ~\underline{\underline{g_{k}}}~ \sum_{k_{\omega}}\frac{1}{\partial_{k}\omega_{k}|_{k=k_{\omega}}}\delta\pc{k-k_{\omega}}}.
 \end{align}

 \textit{\textbf{1d Bath}}:  For one dimensional $k$-space, one can perform the integral along the radial direction (i.e. along the particular coordinate of the k space) by employing the zeroth order tensor (scalar)
 \begin{align}
\underline{\underline{g_{k}}}=(1/\hbar^2)\pr{V_{k}^{2}  k^2}.
\end{align}
To cover the entire $k$-space in this case, one has to count each mode twice. We therefore get
\begin{align}
\underline{\underline{J_{1}}}(\omega)=&\frac{2V}{(2\pi)}\int_{0}^\infty  dk\times\nonumber\\
&\pc{\hbar ~\underline{\underline{g_{k}}}~ \sum_{k_{\omega}}\frac{1}{\partial_{k}\omega_{k}|_{k=k_{\omega}}}\delta\pc{k-k_{\omega}}}.
 \end{align}
 
By inserting the expression for  $ V_{k}$ from Eq.~\eqref{eq:CouplingCoeff}, we perform integrations in all the cases for each of the tensor components.  Due to the symmetry of the $k$ space, the integrals of the off-diagonal elements are zero. The final formula for the spectral density tensor in  $d$-dimensions is
 \begin{align}
\underline{\underline{J_{\rm d}}}(\omega)=\frac{ \mathcal{J}_{\rm d}(\omega)}{d} \underline{\underline{I} }~_{\rm d\times d},
 \end{align}
where $ \underline{\underline{I} }~_{\rm d\times d}$  is the identity matrix and the scalar function $ \mathcal{J}_{\rm d}(\omega)$ in dimension $d$ is 
\begin{align}
&\mathcal{J}_{\rm d}(\omega)=\pc{\frac{S_{\rm d} \pc{\sqrt{2}}^{d}(\eta_{\rm d})^{2}(\Lambda_{\rm d})^{d+2}}{(2\pi)^d}}\times\nonumber\\
&\pc{\frac{\left[\left(\frac{m_{\rm B}}{\pr{g_{\rm B, d}}^{\pr{\frac{d}{d+2}}}n_{0,\rm d}}\right)\left(\sqrt{\frac{\omega ^2}{(\Lambda_{\rm d}) ^2}+1}-1\right)\right]^{(\frac{d+2}{2})}}{\left(\sqrt{\frac{\omega ^2}{(\Lambda_{\rm d})^2}+1}\right) }}.
\label{}
\end{align}
Here for $ d=1, ~2$ and  $3$ we have $S_{1}=2,~ S_{2}=2\pi$ and $S_{3}=4\pi$, respectively. Moreover, we have introduced the characteristic frequency $ \Lambda_{\rm d}=(g_{\rm B,d}n_{0,{\rm d}})/\hbar$.  We also write the impurity-boson coupling  in the units of the boson-boson coupling as $\eta_{\rm d}=(g_{\rm IB,d}/g_{\rm B,d})$ (see main text for $d$-dependence of these quantities). This justifies the formula for the spectral density tensor for $d-$ dimensional  bath, used in Section III.

\section{Vectorial equation of motion}
\label{App:BB}
In this Appendix, we derive the equations of motion for the coordinates of impurity. We restrict the discussion to the three-dimensional case. We start by combining  Eq. \eqref{eq:EOMPositionImpurity} and  Eq. \eqref{eq:EOMMomentumImpurity} of the main text to get the  vectorial equation 
\begin{align}
\underline{\ddot X}(t)+\Omega^{2}\underline{ X}(t)=-\frac{i\hbar}{m_{\rm I}} \sum_{k}\underline{g_{k}}(b_{k}(t)- b_{k}^\dagger(t) ).
 \label{AppBEq1}
 \end{align}
The time-dependent bosonic annihilation and creation operators of the Bogoliubov modes can be extracted from the first-order linear inhomogeneous equations Eq. \eqref{eq:EOMBogoliubovOperator} and Eq. \eqref{eq:EOMBogoliubovOperatorTC}.
\begin{align}
&b_{k}(t)=b_{k}e^{-i\omega_{k}t}+h^{-}_{k}(t),\nonumber\\
&b^{\dagger}_{k}(t)=b^{\dagger}_{k}e^{+i\omega_{k}t}+h^{+}_{k}(t). 
\label{AppBEq2}
\end{align}
Here, $h^{-}_{k}(t)$  and  $ h^{+}_{k}(t)$ represent particular solutions of the following two inhomogeneous differential  equations
\begin{align}
\dot b_{k}(t)+i\omega_{k}b_{k}(t)=-\underline{g_{k}}^{T} \underline{X}(t),
\label{AppBEq3}
 \end{align}
 \begin{align}
\dot b_{k}^\dagger(t)-i\omega_{k}b_{k}^\dagger(t)=-\underline{g_{k}}^{T} \underline{X}(t).
\label{AppBEq4}
 \end{align}
It is obvious that an excitation in the Bogoliubov mode of momentum $k$ depends on all coordinates of the impurity.  Employing the technique of Green's function we construct the solutions corresponding to each direction (cf. \cite{2017Lampo}) and using the superposition principle, we get
\begin{align}
h_{k}^{\pm}(t)=\int_{0}^{t}(1/2)e^{\mp i\omega t}\underline{g_{k}}^{T} \underline{X}(s)ds.
\label{AppAEq5}
 \end{align}
We insert these expressions into Eq. \eqref{AppBEq2} and substitute the result into the right-hand side of Eq. \eqref {AppBEq1}, to obtain the equation of motion of the impurity coordinates
\begin{align}
&\underline{\ddot X}(t)+\Omega^{2}\underline{ X}(t)-\frac{\hbar}{m_{\rm I}}\sum_{k}\underline{\underline{g_{k}}}\int_{0}^{t}\underline{X}(s)\sin[\omega_{k}(t-s)]ds\nonumber\\
&=(1/m_{\rm I})\underline{B}(t),
 \label{AppAEq1}
 \end{align}
where the vectorial Brownian stochastic force represents 
\begin{align}
 \underline{B}(t)= \sum_{k}i\hbar \underline{g_{k}}(b^{\dagger}_{k}e^{+i\omega_{k}t}-b_{k}e^{-i\omega_{k}t}).
\end{align}
By using Eq. \eqref{eq:NoiseKernal} and Eq. \eqref{eq:Spectral DensityDiscrete1}, we further write the above expression in terms of the noise tensor:
\begin{align}
&\underline{\ddot X}(t)+\Omega^{2}\underline{ X}(t)-\frac{1}{m_{\rm I}}\int_{0}^{t}\underline{\underline{\lambda}}(t-s)\underline{X}(s)ds\nonumber\\
&=(1/m_{\rm I})\underline{B}(t).
\end{align}
Since noise and damping kernel are related by
\begin{align}
&-\frac{1}{m_{\rm I}}\int_{0}^{t}\underline{\underline{\lambda}}(t-s)\underline{X}(s)ds=\int_{0}^{t}\underline{\underline{\dot{\Gamma}}}(t-s)\underline{X}(s)ds\nonumber\\
&=\partial_{t}\int_{0}^{t}\underline{\underline{\Gamma}}(t-s)\underline{X}(s)ds-\underline{\underline{\Gamma}}(0)\underline{X}(s),
 \label{}
 \end{align}
we finally arrive at
\begin{align}
&\underline{\ddot X}(t)+\tilde{\Omega}^2\underline{\underline{I}}~\underline{ X}(t)-\partial_{t}\int_{0}^{t}\underline{\underline{\Gamma}}(t-s)\underline{X}(s)ds\nonumber\\
&=(1/m_{\rm I})\underline{B}(t),
 \label{}
 \end{align}
where we have  introduced a renormalized frequency of the impurity:
\begin{align}
\tilde{\Omega}^2\underline{\underline{I}}=\Omega^2\underline{\underline{I}}-\underline{\underline{\Gamma}}(0).
 \label{eq:RenormFreq}
 \end{align}
From here on,  we will neglect such frequency re-normalisation contributed by the term $\underline{\underline{\Gamma}}(0)$. This term grows as the interaction strength between the impurity and the bath increases, and could potentially lead to a negative re-normalised frequency for the harmonically trapped impurity. This, in practice, would correspond to effectively having an impurity trapped in an inverse parabolic potential, for which no stable solution in the long time limit exists.  In view of this, as in  \cite{2017Lampo}, 
we make sure that we always consider values of the parameters for which this renormalized frequency is positive. This issue could have equivalently been solved by artificially introducing a counter-term in the Hamiltonian that would guarantee positivity of the Hamiltonian and translational invariance, but we prefer to use the physical Hamiltonian that we obtained directly from the Hamiltonian describing the original Bose polaron.  

\section{Expressions of the position and momentum variances for the generalized Langevin equations}
\label{App:CC}
The quantum covariance matrix is defined as \cite{ferraro2005gaussian}
\begin{align}
\sigma_{kl}=\frac{1}{2}\langle\{R_{k},R_{l}\}\rangle-\langle R_{k}\rangle\langle R_{l}\rangle.
 \end{align}
Here the $R_{k}$ represent the quadratures of the motion. For zero mean value, the matrix contains in particular the variances of the impurity position  $\langle x^{2}(t) \rangle$  and  momentum  $\langle p^{2}(t) \rangle$  as  its diagonal terms. Note that we have omitted  the operator notation for  convenience. Assuming that the bath and system variables are initially uncorrelated, we obtain the expressions 
\begin{align}
&\langle x^{2}(t) \rangle=G_{1}^{2}(t)\langle x^{2}(0) \rangle
+G_{2}^{2}(t)\langle \dot x^{2}(0) \rangle+\nonumber\\
&\frac{1}{2 m_{\rm I}^{2}}\int_{0}^{t}du\int_{0}^{t}dv G_{2}(t-u)G_{2}(t-v)\langle\{B(u),B(v)\}\rangle,
\label{eqapp:positionvarience}
 \end{align}
\begin{align}
&\langle p^{2}(t) \rangle=m_{\rm I}^{2}\dot G_{1}^{2}(t)\langle x^{2}(0) \rangle+\dot G_{2}^{2}(t)\langle  p^{2}(0) \rangle+\nonumber\\
&\frac{1}{2}\int_{0}^{t}du\int_{0}^{t}dv \dot G_{2}(t-u)\dot G_{2}(t-v)\langle\{B(u),B(v)\}\rangle.
 \end{align}

The expression for the mean square displacement $ {\rm MSD}$ takes the form
\begin{align}
&{\rm MSD}(t)=(G_{1}(t)-1)^{2}\langle x^{2}(0) \rangle+G_{2}^{2}(t)\langle \dot x^{2}(0) \rangle+\nonumber\\
&\frac{1}{2 m_{\rm I}^{2}}\int_{0}^{t}du\int_{0}^{t}dv G_{2}(t-u) G_{2}(t-v)\langle\{B(u),B(v)\}\rangle.\nonumber\\
 \end{align}
 \section{Validity of Fr\"{o}hlich Hamiltonian}
 \label{App:DD}
 In the main text we use the linear Fr\"{o}hlich  Hamiltonian while discarding  the two-phonon scattering processes. This is based on the assumption that the condensate density in the $d$-dimension, $n_{0,{\rm d}}$, is much larger than the density of the phonons  excited due to the interaction with the impurity \cite{inguscio2016quantum}. As stated in \cite{PhysRevA.76.011605}, an approximated criterion on the coupling parameter $\eta_{\rm d}$ for the Fr\"{o}hlich Hamiltonian to be valid in the $d$-dimension is given by
\begin{align}
\eta_{\rm d}&<\eta_{c,{\rm d}}\nonumber\\
&  =\! \! \pc{\sqrt{\frac{4(2\pi)^d}{S_{\rm d}}}}n_{0,{\rm d}}~\pc{\xi_{\rm d}}^{d}\! =\! \pc{\! \hbar \sqrt{\frac{2(2\pi)^d}{S_{\rm d}}}}\! \! \frac{c_{\rm d}~\pc{\xi_{\rm d}}^{d-1}}{g_{\rm B,d}}\nonumber\\
&=\! \! \pc{\sqrt{\frac{2^{2-d}(2\pi)^d}{S_{\rm d}}}}  \left[ n_{0,\rm d}\right]^{\frac{2-d}{2}}\left[\frac{\pc{\sqrt{\hbar/m_{\rm B}\omega_{\rm d}}}^{3-d}}{ S_{\rm d}  a_{3}}\right]^{\frac{d}{2}}.
 \end{align}
In the above expression we have written the final equality in terms of the three dimensional scattering length $a_{3}$ based on the harmonic confinement of the condensate in the transverse direction (see main text). From the second-to-last inequality one can get the expression for $d=3$
\begin{align}
g_{\rm IB,3}\lesssim2\pi c_{3}\pc{\xi_{3}}^{2} 
 \end{align}
 (where we put $\hbar=1$).
The same bound was reported in \cite{inguscio2016quantum}. On the other hand, for $d=1$, the last equality  leads to the  following scaling for the bound on the critical coupling:
\begin{align}
\eta_{c,{\rm d}}\sim\sqrt{n_{0,1}a_{1}}~; \quad\quad  \text{where}~  a_{1}=\pc{\hbar/m_{\rm B}\omega_{\perp}}/a_{3},
 \end{align}
 stated also in \cite{2017Lampo, Grusdt2017}. Typically for a boson gas made of ${\rm Rb}^{87}$ atoms,  the scattering length $ a_{3}=100a_{0}$ with  $a_{0}$ denoting the Bohr radius \cite{inguscio2016quantum}. Moreover, let the transverse frequencies be $\px{\omega_{\perp},\omega_{z}}=2\pi \times 34~{\rm kHz}$  as in the optical lattice  \cite{Catani2012}. This implies the following numerical bounds on the coupling, depending on the dimension:
 \begin{align}
\eta_{c,{\rm d}}\sim\begin{cases}
 3.7 ,~  & \text{for}~\quad d=1,\\
 4.4,   ~  &\text{for }~ \quad d=2,\\
9.8 , ~&\text{for }~ \quad d=3,\\
\end{cases}
 \label{}
 \end{align}
 which are the values represented as vertical lines in the figures in Figs. \ref{fig:fig2} and \ref{fig:fig6-7}.


\bibliographystyle{apsrev4-1}
\bibliography{References1.bib}

\end{document}